# Newly confirmed and candidate Galactic SNRs uncovered from the AAO/UKST Hα survey

M. Stupar,[1,2] Q. A. Parker,[1,2] M. D. Filipović [3]

[1] Department of Physics, Macquarie University, Sydney 2109, Australia
[2] Anglo-Australian Observatory, P.O. Box 296, Epping, NSW, 1710, Australia
[3] University of Western Sydney, Locked Bag 1797, Penrith South, DC, NSW 1797, Australia

**Abstract**

We present a catalogue of 18 new Galactic supernova remnants (SNRs) uncovered in the optical regime as filamentary emissions and extended nebulosities on images of the Anglo Australian Observatory/United Kingdom Schmidt Telescope (AAO/UKST) H$\alpha$ survey of the southern Galactic plane. Our follow-up spectral observations confirmed classical optical SNR emission lines for these 18 structures via detection of very strong [SII] at 6717 and 6731Å relative to H$\alpha$ ( [SII]/H$\alpha$ > 0.5). Morphologically, 10 of these remnants have coherent, extended arc or shell structures, while the remaining objects are more irregular in form but clearly filamentary in nature, typical of optically detected SNRs.

In 11 cases there was a clear if not complete match between the optical and radio structures with H$\alpha$ filamentary structures registered inside and along the presumed radio borders. Additionally, ROSAT X-ray sources were detected inside the optical/radio borders of 11 of these new remnants and 3 may have an associated pulsar.

The multi-wavelength imaging data and spectroscopy together present strong evidence to confirm identification of 18 new, mostly senile Galactic SNRs. This includes G288.7-6.3, G315.1+2.7 and G332.5-5.6, identified only as possible remnants from preliminary radio observations. We also confirm existence of radio quiet but optically active supernova remnants.

(ISM:) supernova remnants: emission line - surveys (optical)

# 1 Introduction

An up-to-date inventory of Galactic SNRs provides key data for improving our understanding of Galactic star formation history and the Galactic energy balance. Unfortunately, the total number of known SNRs (265, Green 2006) compared to those expected (>1000, Case & Bhattacharya 1998) implies there remains a significant population of undiscovered remnants. This is primarily due to various selection effects, such as highly variable extinction and the wide range of observed properties depending on the SNRs evolutionary state. These hamper detection of both the youngest and oldest SNRs of large angular size which are usually highly fragmented. In the modern era, identification and observation of most Galactic SNRs has been predicated by radio detections where their non-thermal spectra and typically coherent radio shell structures and low linear polarization are easy to recognise. Radio observations substantially mitigates the problems of dust which make optical detection so difficult. Indeed, due to the high concentration of dust around the Galactic plane, prior to this work only a small fraction of known Galactic SNRs, as detected over a wide range of radio frequencies, have matching optical detections. Green's on-line SNR catalogue (Green 2006) of 265 Galactic SNRs, the most complete compilation currently available, report optical detection in only ~17% of SNRs. However, radio surveys are not very sensitive to extremely evolved SNRs.

The second major tool in SNR investigations today are X-ray observations via satellites which enables estimation of key parameters such as remnant age, explosion energy and the ambient interstellar density (e.g. Mavromatakis et al. 2004). Here it has been shown that SNR X-ray morphology does not follow the typical radio shell structure. SNRs exhibit more mixed X-ray morphology, typically with filled-centres and with both thermal and non-thermal emission (Seward 1990).

Historically, many Galactic SNRs were first detected in the optical via the early wide-field photographic surveys (e.g. Rodgers, Campbell & Whiteoak 1960), particulary in the line of H$\alpha$. The optical morphology, often filamentary in nature, was seen to follow the outline of the radio shell revealed in subsequent radio observations. Also common was the presence of highly fragmented, narrow, optical filaments or emission clouds of irregular form both inside, outside or on the given radio boundaries.

Over the last few years a series of papers by an active Greek group (e.g. Boumis et al. 2005, 2007) have been uncovering new, optical SNR detections by targeting many known northern remnants via selected, deep, narrow-band imaging and spectroscopy, reinforcing the value of optical coverage. However, individual narrow optical filaments, elongated arc structures or irregular nebulous clouds cannot be accepted as part of an SNR until other supporting evidence is available, such as provided by non-thermal radio signatures or optical spectra which verify remnant shocked gaseous emission.

Optical spectroscopic signatures typical of SNRs and of shocked gas include strong forbidden line emission such as [SII] 6717/6731Å, [NII] 6548/6584Å, [OI] 6300/6364Å, [OIII]4959/5007Å and [OII] 3727Å. The [SII] lines should be strong and particularly high ratios of [SII]/$\alpha$ which enable a clear distinction between HII regions, most planetary nebulae and SNRs where the detected spectral lines are similar. A value of [SII]/H$\alpha > 0.5$ is often used as the basis for separating SNRs from HII regions and PNs. In fact, Mathewson & Clarke (1973) adopted ~0.7, while Daltabuit, D'Odorico & Sabbadin (1976) suggested ~0.5. Fesen, Blair & Kirshner (1985) accepted [SII]/H$\alpha > 0.5$ coupled with the presence of strong [OI] and [OII] lines (but only where the spectral resolution permits proper [OI] sky subtraction). The value ([SII]/H$\alpha > 0.5$) has been used as a key spectral discriminator in this work, together with the supporting presence of Balmer, [NII], [OII] or [OIII] lines.

The AAO/UKST SuperCOSMOS H$\alpha$ survey (SHS hereafter; Parker et al. 2005) with its combination of arcsecond resolution, 5 Rayleigh sensitivity and 4000 sq. degree coverage, now offers a powerful, new opportunity to revisit the regime of optical SNR detection and discovery across the entire southern Galactic plane. The SHS comprises 233 overlapping fields on 4$°$ field centres (Parker & Phillipps 1998). Every survey pointing was exposed for 3 hours, accompanied by a matching 15 minute broad-band (5900-6900Å) short red (SR) exposure. The SHS has already resulted in significant discoveries of planetary nebulae (e.g. Parker et al. 2006; Miszalski et al. 2008), Wolf Rayet stars (Morgan, Parker & Russeil 2001;

Parker & Morgan 2003), HII regions and filamentary nebulosities (Walker, Zealey & Parker 2001) and is a rich source of emission sources of every kind. Detailed confirmations of 3 SNR candidates based on our SHS discoveries have already been reported by (e.g. Stupar, Parker & Filipović 2007a,b; Stupar et al. 2007c).

For all currently known Galactic SNRs there are only 15 with $|b|$ >5 degrees and their distribution is quite asymmetric. Only 4 are in the Southern Hemisphere, while the rest are above $\pm 5$ degrees (Green2006). Although this asymmetry may be related to Gould's belt (Stothers & Frogel 1974), there is no evidence of the same asymmetry for low galactic latitude SNRs, which are not statistically different (Green2006). The SHS extends to $|b|\sim 10-12$ degrees and provides a fresh opportunity to search for evolved remnants to higher Southern latitudes in the optical.

Here we report results of a systematic SHS search, undertaken to improve the detection of both known and new Galactic SNRs across the southern Galactic plane from a non-radio perspective. We present results of our extensive follow-up spectral observations of new candidates obtained during five runs on the 1.9m telescope of the South African Astronomical Observatory and 2.3m telescope of Mount Stromlo and Siding Spring Observatory between 2004-2006. These observations provided confirmation of the 18 new Galactic SNRs in this paper. We provide compelling evidence for the veracity of these new SNRs through the combination of diagnostic optical spectroscopic signatures and likely radio, X-ray and pulsar connections with many of the H$\alpha$ morphological structures identified from the SHS. We conclude that most of these new remnants are senile and probably in the dissipation or merging phase with the ISM. This has made previous identification from the extant radio surveys problematic, due to high levels of fragmentation, but more obvious once the radio survey data is superimposed on the SHS images. We have now added ~8% to the total number of remnants recognized in the Galaxy. Many additional candidates still require spectroscopic follow-up and further confirmations will be presented in subsequent papers.

## 2 Identification of new Galactic SNRs in the AAO/UKST H$\alpha$ Survey (SHS)

Our large-scale, systematic search for new Galactic SNRs was based initially on detection of new, coherent, optical emission structures identified in the SHS (e.g. Stupar, Parker & Filipović 2007a,b; Stupar et al. 2007c).

Although the SHS survey is an H$\alpha$ survey, the interference filter used has a central wavelength at $6950\pm 25$Å and FHWM of $70\pm 3$Å (Parker & Bland-Hawthorn 1998) so both [NII] emission lines at 6548 and 6584Å are included. These contributions can only be effectively deconvolved via spectroscopy. However, this exceptional monolithic H$\alpha$ filter avoids problems with previous segmented filters (Meaburn 1980) and provides a very uniform survey once flat-fielded (Parker et al. 2005). Using the SuperCOSMOS measuring machine at the Royal Observatory Edinburgh (Miller et al. 1992), the matching H$\alpha$ and SR images were transformed into digital form and have been available on-line since 2003[1]. For practical access reasons, only 900 square arc minutes of pixel data as FITS images can be downloaded at a time. The pixel data for each survey field is about 2GB. Every downloaded fits image has accurate in-built astrometry (world co-ordinate system).

### 2.1 Examination of SHS blocked-down data

Downloading $30\times 30$ arcmin full resolution images of the entire 4,000 square degrees of the H$\alpha$ survey was impractical so, an initial search for large, coherent H$\alpha$ emission structures was undertaken using the $16\times 16$ blocked FITS images of all survey fields which are also on-line. An advantage of the blocked-down data is that large-scale, low surface brightness features are enhanced. However, the resolution is relatively low (~11 arcseconds) but still compares favorably with the 48 arcsecond resolution of the SHASSA on-line H$\alpha$ survey (Gaustad et al. 2001). All 233 blocked-down SHS fields were carefully examined to identify about 60 new candidate emission structures. The RA/DEC coordinates of these candidate regions were then used to extract full resolution SHS images of appropriate size for detailed study. This strategy enabled all 233 survey fields to be examined efficiently and provided many decent candidates but there is a problem. Smaller scale, fractured and fine filamentary and weak emission features can easily be overlooked in the blocked-down images yet these are some of the best optical manifestations of evolved remnants.

---
[1] http://www-wfau.roe.ac.uk/sss/halpha/

### 2.2 Examination of original H$\alpha$ survey films

Due to resolution limitations of the blocked SHS data coupled with the download field size restrictions of the high resolution images, examination of the original survey fields was undertaken at the Plate Library of the Royal Observatory Edinburgh (ROE) in August 2004 where the original H$\alpha$ and counterpart SR films are archived. 29 further small scale filaments and diffuse emission candidates (see example in Stupar, Parker & Filipović 2007a) were found in addition to the ~60 candidates uncovered from the blocked-down FITS images. The visual search of the original films did not cover the regions around the Galactic centre or Vela SNR as both areas are exceptionally rich in emission filaments which requires much more detailed examination.

Table 1. The main characteristics of the newly discovered SNR candidates

| Standard ID | R.A. J2000.0 | $\delta$ J2000.0 | Extent (arcmin) | [NII]/H$\alpha$ (average) | [SII]/H$\alpha$ (average) | Average [SII] 6717/6731 | Radio | X-ray | Pulsar | Remarks |
|---|---|---|---|---|---|---|---|---|---|---|
| G253.0+2.6 | 08 25 54 | -33 26 09 | 6 | 1.07 | 1.16 | 1.11 | Y? | N | N | |
| G243.9+9.8 | 08 28 58 | -21 57 43 | 16 | 0.65 | 0.89 | 1.55 | Y | N | N | Brandt 152 |
| G283.7-3.8 | 10 05 23 | -60 17 07 | 120 | 1.38 | 1.47 | 1.33 | N | Y | N | |
| G288.7-6.3 | 10 26 40 | -64 41 13 | 2.5 | 0.46 | 0.54 | 0.93 | Y | Y? | N | Radio SNR candidate |
| G281.9+8.7 | 10 41 04 | -48 48 49 | 150 | 1.39 | 3.59 | 1.40 | N | Y | N | |
| G288.3+4.8 | 11 07 19 | -55 05 00 | 5,15 | 0.62 | 1.46 | 1.37 | N | Y | N | |
| G289.7+5.1 | 11 17 02 | -55 17 11 | 7,15,22 | 0.60 | 1.38 | 1.40 | Y | Y | N | |
| G289.2+7.1 | 11 18 35 | -53 18 01 | 30 | 0.71 | 1.80 | 1.29 | N | Y? | N | |
| G292.9+4.4 | 11 41 42 | -57 20 00 | 100 | 0.65 | 0.97 | 1.40 | Y | Y? | N | RCW 59 |
| G303.6+5.3 | 12 56 31 | -57 30 21 | 240 | 0.58 | 0.74 | 1.42 | N | Y | Y | |
| G306.7+0.5 | 13 23 35 | -62 0541 | 25 | 0.55 | 0.85 | 1.36 | Y? | Y | Y | |
| G315.1+2.7[*] | 14 33 25 | -57 35 30 | 11 | 1.42 | 1.13 | 1.41 | Y | Y? | N | |
| G332.4+0.6 | 16 13 01 | -50 18 59 | 30 | 1.12 | 1.02 | 1.20 | N | N | Y? | |
| G343.4+3.6 | 16 43 03 | -40 37 44 | 4,7 | 0.81 | 0.59 | 1.35 | Y | N | N | |
| G332.5-5.6[*] | 16 42 17 | -54 28 33 | 30 | 2.42 | 1.92 | 1.23 | Y | Y | N | |
| G329.9-7.8 | 16 43 14 | -58 00 15 | 7 | 1.77 | 1.18 | 1.27 | N | N | N | |
| G348.2-1.8 | 17 18 49 | -39 51 26 | 5,8 | 0.94 | 0.82 | 1.42 | Y | N | N | |
| G18.7-2.2 | 18 33 07 | -13 38 52 | 30(120?) | 0.72 | 1.33 | 1.34 | Y? | N? | N? | |

[*] Indicates the 2 new confirmed remnants which have already been published separately.

During scrutiny of the SHS it was noticed that some of the brighter components of candidate SNRs detectable with existing broad-band optical surveys are already in the literature as nebulosities of unknown nature or as HII regions. Many of these putative HII regions, identified by their morphological structure on broad band images have no spectra to verify their true nature. The SHS sensitivity and resolution reveals far more extensive and distinct features of these regions which brings in to question their original HII identification.

For the first time, we have also uncovered optical filaments and diffuse emissions associated with ~30 known Galactic SNRs previously detected solely via their radio signatures (Stupar, Parker & Filipović 2007b).

### 2.3 Checks against catalogues of existing nebulous objects

Previously identified Galactic HII regions and evolved planetary nebulae can appear similar to SNRs in their optical morphological structure so new remnant candidates were first checked against the SIMBAD[2] database of catalogued nebulae and against the extensive list of ~1250 new `MASH' PNe also discovered from the SHS (Parker et al. 2006; Miszalski et al. 2008). A further check was done against Wolf-Rayet stars with known nebulosities. These stellar outflows sweep up the surrounding material creating a nebulae or interstellar bubble which can appear similar to an SNR shell. In the optical regime, like SNRs, these structures can be recognized in H$\alpha$ and [OIII] light (Chu, Treffers & Kwittek 1983). A further check against entries in the 7$^{th}$ Catalogue of Galactic Wolf-Rayet stars (van der Hucht 2001) was also done to eliminate candidates that are actually just newly identified nebulosities associated with previously bare WR stars, now made possible with the SHS.

### 2.4 Comparison with extant radio data

The most important comparison is with possible radio detections, whether previously associated with a candidate remnant or not. Data for four radio surveys was examined~ the PMN at 4850 MHz (Condon, Griffith & Wright 1993), the SUMSS at 843 MHz (Cram, Green & Bock 1998), the Parkes 2400 MHz survey of the Southern Galactic Plane (Duncan et al. 1995) and the northern NVSS 1.4 GHz (up to $\delta = -40°$) continuum survey (Condon et al. 1998). These data were used to establish any possible connection between radio structures and our newly discovered optical SNR candidates.

Possible radio-optical matches were checked via contouring of the best radio flux images and overlaying these on a grayscale of the H$\alpha$ images. In some cases an excellent match was found between the new optical filamentary structures and the radio contour data (see Stupar, Parker & Filipović 2007a). These present excellent candidates for further confirmatory, multi-frequency follow-up.

### 3 Optical spectroscopic observations

The 5 spectroscopic runs awarded were insufficient to observe all of our SNR candidates due to poor weather. Remaining candidates will form the basis of a subsequent publication once follow-up is available. Two runs were at the 1.9m SAAO telescope (in July 2004 and February 2006) and three runs at the 2.3m MSSSO (in June 2004 and 2005 and August 2005). After the spectra were obtained and reduced, the ratios of key emission lines were used in diagnostic plots to aid classification as new SNRs or other object types accordingly.

For primary object classification, low resolution SAAO 1.9m spectra (~7Å between 3500 and 7500Å) were used, and medium resolution MSSSO 2.3m blue (~2.2Å between 3500 and 5400Å) and red (~1.1Å between 6100 and 6900Å) spectra were used. If the low resolution spectra showed promising SNR lines and ratios, the same object was re-observed with a higher dispersion red grating to cover the H$\alpha$ and [SII] lines for improved radial velocities.

From the observing runs, 18 SNR candidates were confirmed following the line ratio diagnostic criteria of Fesen, Blair & Kirshner 1985 and results are presented here. For ~30 candidates their spectra did not confirm shocked material and the majority of these nebulae are previously uncatalogued HII regions which will be reported in a separate publication.

All spectroscopic runs suffered non-photometric nights so proper extinction estimates could not be calculated directly from the spectra. Hence, for the relative strength of spectral lines (and their ratio), $F(\lambda)$ instead of extinction corrected $I(\lambda)$ was used.

For data reduction standard IRAF slit-spectra routines were used. For flux calibration standard spectrophotometric stars were observed each night when possible from the lists of ham92,ham94. The Starlink FIGARO package BCLEAN was used for cosmic ray rejection while extraction of 1 D spectra was facilitated via supplementary IRAF and perl scripts developed by Brent Miszalski[3].

Table 1 presents the main characteristics for all 18 SNR candidates. The first column gives the standard remnant ID as truncated Galactic coordinates. For Galactic SNRs the attributed name is usually formed from the Galactic coordinates of the remnant's centre (usually in the form of a shell) where the progenitor star was probably located before the supernova explosion. Due to the irregular structure of these newly uncovered filaments/clouds it was often not possible to properly define the overall shape of the SNR from the optical data nor to estimate the centre of the observed structures. Under these circumstances, the SNR name was taken from the Galactic coordinates of the slit position used to obtain the primary optical spectrum, usually taken across the most prominent optical structure. Exceptions are G288.7-6.3, G292.9+4.5 (RCW 59), G315.1+2.7 and G332.5-5.6 previously known as SNR candidates or HII regions (RCW 59). In Table 1 we also include the ratio of the [SII]/H$\alpha$ lines as this is the most important spectroscopic diagnostic to separate SNRs from HII regions and most planetary nebulae. If the optical imagery and spectral confirmation is supported with any radio, X-ray or pulsar detection in the vicinity this is also noted.

---

[2] http://simbad.u-strasbg.fr/

[3] http://www.aao.gov.au/local/www/brent/pndr/

For determining the uncertainties in the flux estimates the IRAF task *splot* was used to fit Gaussian profiles to provide both fluxes and a standard deviation $\sigma$ which was also used as an average measure of the Gaussian line profile of all brightest lines. The flux error for the brightest lines with the SAAO 300 lines/mm grating was between 5 and 10%, while for the medium resolution 1200 lines/mm red grating it was around 20%. This is not unexpected due to the generally lower S/N obtained for the higher resolution spectra. For the MSSSO 2.3m blue grating (600 lines/mm) this error is between 10 and 20% and for the red grating (1200 lines/mm) it is between 15 and 25%. Due to the lower detector sensitivities in the the blue the spectral region between 3500 and 4000Å has low S/N. Unfortunately, the valuable diagnostic [OII] 3727Å emission doublet is located here and flux calibration of this particular line was often very uncertain.

### 3.1 Other detected SNR diagnostic emission lines

The most prominent optical spectral diagnostic for SNRs is the ratio of [SII]/H$\alpha$ being > 0.5 which indicates the presence of shocks. Shocks arise from other astrophysical phenomena but high values of this ratio are strongly indicative of an SNR. Only 2 of the new SNRs have this value <0.7 while for 12 new remnants this value is ≥ 1 for at least one slit position. There are other line ratios that are important in supporting SNR classification such as the ratio of the [OIII] lines (4959+5007)/4363 (an electron gas temperature indicator) having a value < 30 (Fesen, Blair & Kirshner 1985) which implies electron gas temperatures in excess of 20,000K, values never seen even in the hottest HII regions. Our observations are of faint new remnants with modest spectral S/N so these very faint lines are not detected and play no part in our classification.

The oxygen [OI] lines at 6300 and 6364Å are present and often prominent in most SNR candidate spectra. This is not the case for HII regions or PNe. Unfortunately, these lines are also strong in the night-sky spectrum and are susceptible to imperfect sky-subtraction as the nebulae usually extend quite a way across the slit. Even with careful sky subtraction the expected 6300/6364Å lock-step ratio of 3:1 was often not achieved due to the modest spectral resolution so the flux values for these lines are not reliable. However, [OI] detection is often verified from careful examination of the 2-D spectral image. Other typical SNR emission lines are sometimes seen in our sample, such as [NeIII] + H at 3970Å or He + H at 3888Å (Fesen, Blair & Kirshner 1985). The Balmer lines (in different strengths) are always present in the observed SNR candidate spectra but are not recorded for brevity and detection in the blue depends on extinction.

### 3.2 The new SNR Catalogue

We present details and H$\alpha$ images of all 18 new SNR candidates that have passed the various selection criteria. For objects where a clear radio connection at 843 MHz (SUMSS radio survey), 1400 MHz (VLA NVSS radio survey), 2400 MHz (Parkes radio survey) or 4850 MHz (PMN radio survey) has been found, the radio contours were overlaid on the optical image as support of the SNR classification. In a few cases radio detections at several frequencies was noticed, offering the potential to measure a spectral index to further bolster the SNR classification (e.g G332.5-5.6), but in most instances only single frequency detections are possible from the existing radio surveys. Targeted multi-frequency follow-up of these new SNRs is required to reveal their non-thermal nature.

Any association of a candidate with an X-ray source or possible pulsar in close proximity is also indicated in the catalogue even though it is often hard to establish a definite connection between these objects and our candidates. Some images also indicate the slit position(s) for the spectroscopic follow-up. Candidates are presented according to increasing RA as in Table 1.

**G253.0+2.6**

This candidate was first recognized as a discrete group of H$\alpha$ filaments (Fig. 1) where the largest arcuate filament is about 6 arcmin at maximum extent and was discovered on SHS field HA528. Due to the large overlap that exists between adjacent SHS survey fields, this structure was also clearly visible on survey field HA602, confirming its veracity. Two SAAO 1.9m low resolution spectra were obtained along the major filament sampled about 2.5 arcmin apart exhibiting typical SNR characteristics with strong [NII] and [SII] lines.

The PMN 4850 MHz radio survey data in the region shows some weak detection though precise details are not possible due to the low survey resolution (FWHM of the 43 m dish is ~7 arcmin). No detection was found in the VLA NVSS 1.4 GHz survey (which covers up to -40° in declination), while the SUMSS 843 MHz does not cover this area of sky. Further radio observations are warranted.

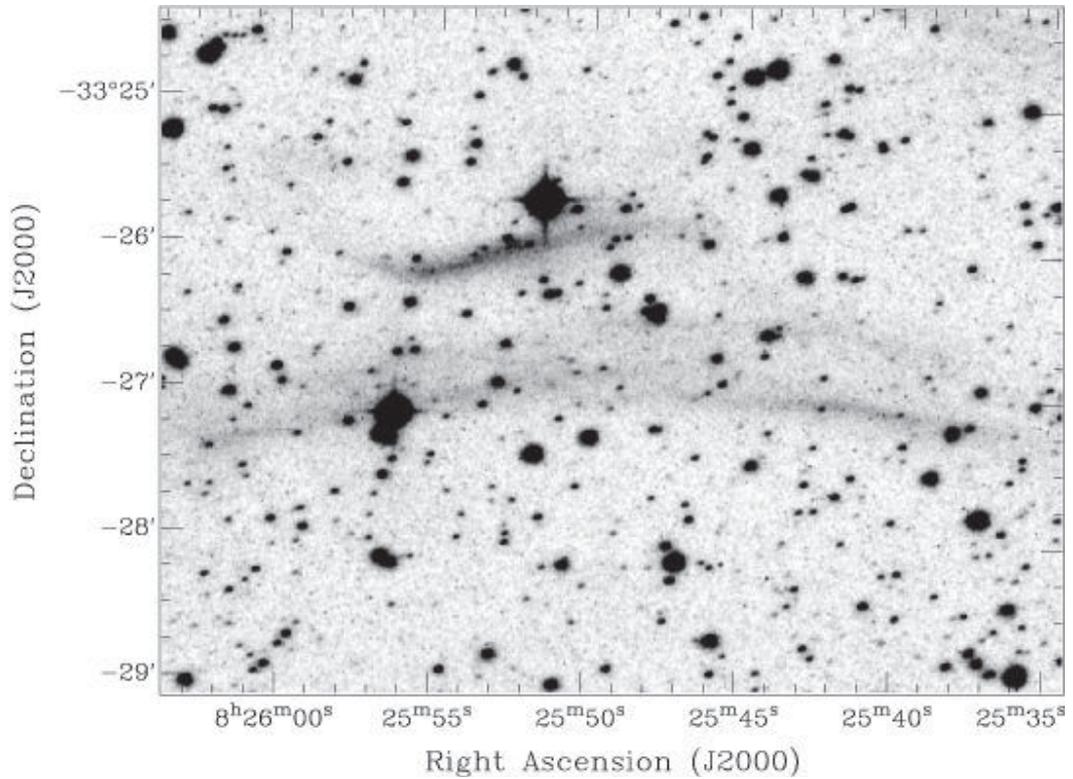

Figure 1. G253.0+2.6. The group of filaments concentrated around RA=08$^h$ 25$^m$ 50$^s$ $\delta$=-33° 25' 30". The extent of the largest arcuate filament is ~6 arcmin. Fragmented filaments occupy an area of ~8 arcmin.

**G243.9+9.8**

This is a large, prominent, arcuate nebulosity (Fig. 2) about 16 arcmin at maximum extent discovered on SHS field HA761. Only the brightest component was previously identified as Brand 152, a nebula of unknown nature (Brand, Blitz & Wouterloot 1986). The new SHS imaging reveals further and much more extensive structural information that indicated it might be an SNR candidate worthy of spectroscopic follow-up. There is a strong morphological similarity between this object and known Galactic SNR IC 443.

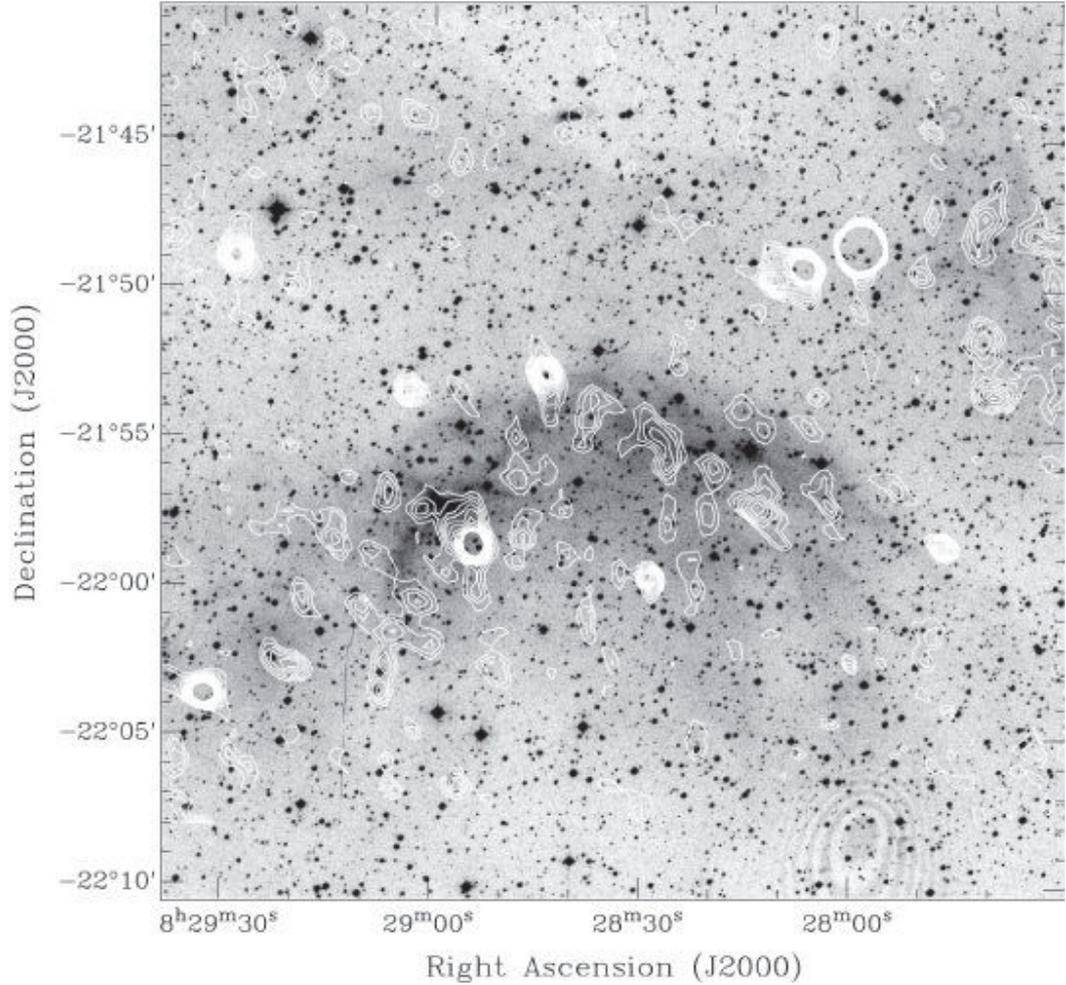

Figure 2. 30×30 arcmin field around G243.9+9.8. The Hα image is overlaid with the NVSS 1.4 GHz continuum survey contours. The strongest contours are 0.006 Jy beam$^{-1}$. Although highly fragmented there is a clear excess of radio power following the main optical arc structure. The faint elliptical forms in the lower right corner are interference patterns (Newton's rings) due to the SuperCOSMOS glass platten used to sandwhich the SHS Tech-Pan films in place during scanning.

From SIMBAD there are no pulsar, X-ray or radio sources catalogued in the immediate vicinity. However, a separate check against existing radio survey data reveals a match of radio contours from the VLA NVSS 1.4 GHz with the optical image (Fig. 2). The radio contours are fractured but they clearly follow the optical arcuate structure. This is an excellent example of how the existing fractured radio data makes it hard to pick up the candidate from the radio data a priori but how easy it is to make the connection once combined with the new Hα imaging. This object is situated at the extreme N-W edge of the Gum nebula. The similarity of certain structural aspects and close proximity to this enormous structure cannot rule out that this candidate is connected with this large nebula. However, projection of this new SNR across this part of the Gum nebula is also possible. Support for this conclusion comes from the work of bbw86 where this object is also taken to be separate from the Gum nebula. Follow-up spectroscopy shows strong lines of [SII] with [SII]/Hα ~0.89 and very strong Hβ and [OII] at 3727Å.

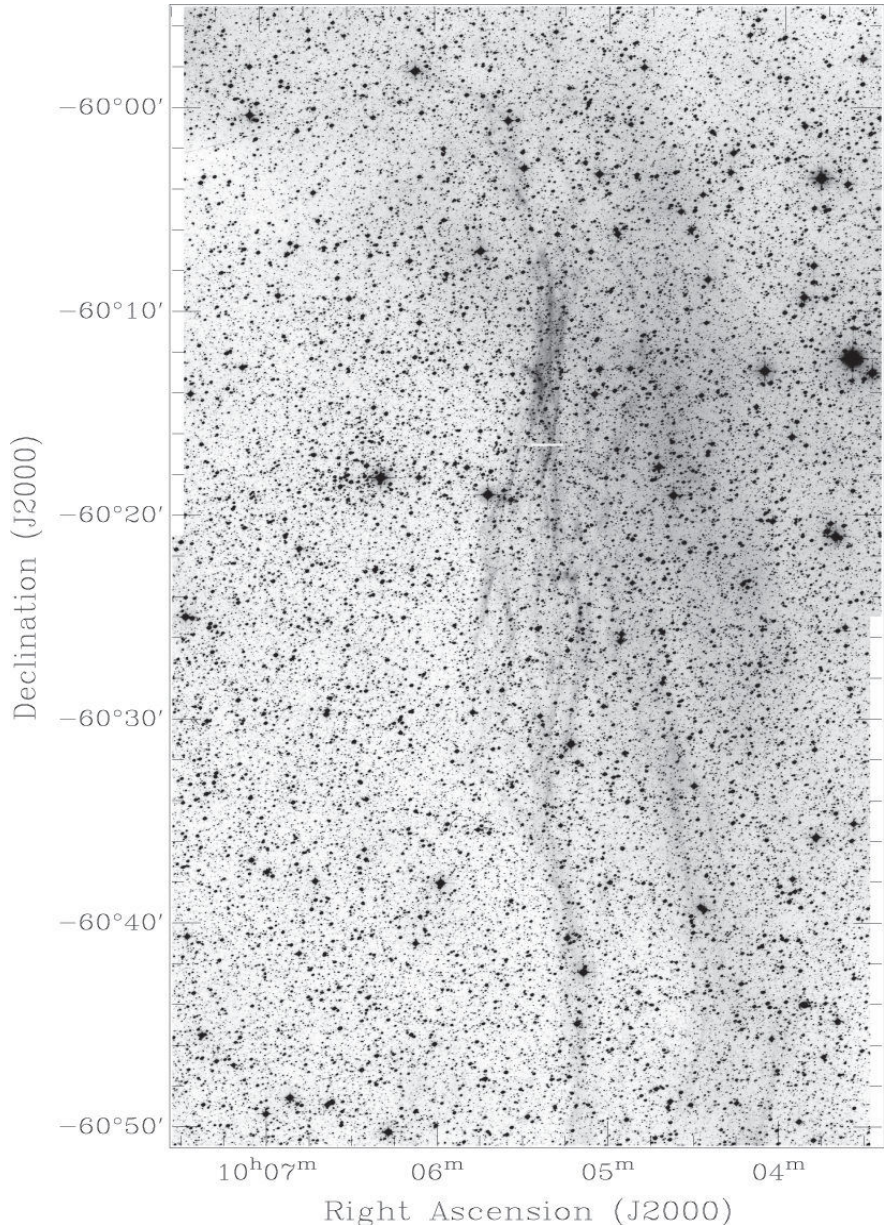

Figure 3. G283.7-3.8. H$\alpha$ high resolution mosaic image (~1°×0.5°) of the most prominent part of G283.7-3.8 showing parallel filaments with maximal extent of 2°. Three SAAO spectra (two low resolution and one with medium resolution of 1200 lines/mm) confirm that G283.7-3.8 as a likely supernova remnant.

**G283.7-3.8**

This object comprises a significant group of parallel filaments (Fig. 3) with a maximum extent of 2° and was discovered on SHS field HA174. There is a Wolf-Rayet star WR 17 (HD 88500) vanderh01 at R.A.=$10^h 10^m 31.92^s$ and $\delta$=-60° 38′ 42.40″ (J2000.0) about 40 arcmin east from the filaments. The high quality SHS imaging reveals further structural detail indicating the filaments are unrelated and different in type from the morphological structure of the WR star's nebulae so we do not consider G283.7-3.8 to be an outer, very distant part of the WR nebula but a new SNR candidate. Spectroscopy across three of these filaments confirms this, with [SII]/H$\alpha$ giving a consistent, very high ratio of between 1.2 and 1.8 despite the distance between the slit separations of ~1°. Filaments were observed at the same position with both the low (300 lines/mm) and high (1200 lines/mm) resolution gratings with the red spectrum ranging from 6150 to 6900Å. The spectra exhibit characteristic SNR optical emission lines including extremely strong [OII] at 3727Å. The [OIII] lines at 4959 and 5007Å are also present.

SIMBAD did not reveal any catalogued pulsar, X-ray or radio source in the vicinity and inspection of the PMN and SUMSS surveys did not give any significant radio detection though this may be due to radio dish insensitivity to such large structures.

**G288.7-6.3**

This optical H$\alpha$ nebulosity (see Fig. 4), about 2.5 arcmin at maximum extent, was discovered on SHS field HA132. This nebulous patch matches the N-W part (see Fig. 5) of the candidate SNR G288.7-6.3 (Duncan et al. 1995) so it is reasonable to conclude that this nebulosity may represent the first detected optical counterpart of this candidate SNR. SIMBAD did not confirm the existence of any catalogued objects at the position of the nebulosity, but did return 9 X-ray sources inside the larger candidate G288.7-6.3 (which has an overall radio extent of ~2°).

The medium dispersion 2.3m spectrum taken shows reasonable SNR spectral characteristics with the ratio [SII]/H$\alpha$ = 0.54. Such values have been known as upper values for some HII regions (see discussion in Fesen, Blair & Kirshner 1985). The ratio of [NII]/H$\alpha$ is also low which although not ruling out an HII region identification and the value is at the lower level found for SNRs, but is still in accordance with different abundances of nitrogen inside the Galaxy. In the 2.3m blue spectrum however extremely strong [OII] emission at 3727Å is seen together with prominent [OIII] at 5007Å (359 and 91 against H$\beta$ is seen), which effectively eliminates a HII region but not a PN.

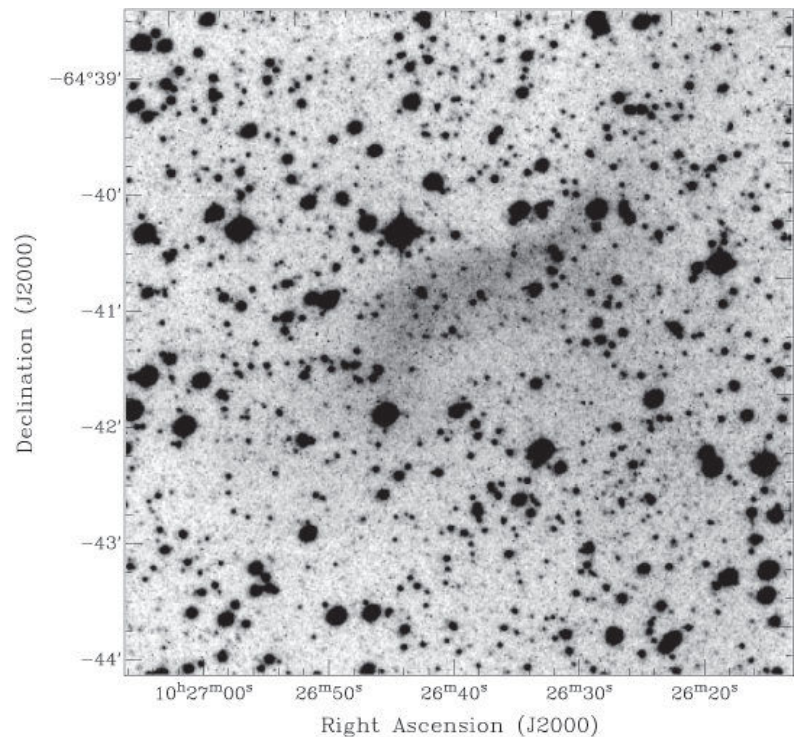

Figure 4. A small H$\alpha$ nebula marked as G288.7-6.3, approximately 2.5 arcmin at maximum extent.

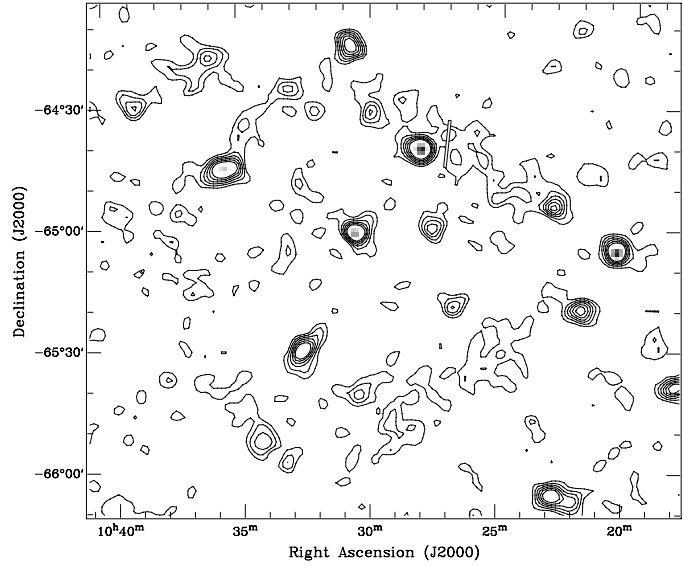

Figure 5. G288.7-6.3 candidate SNR from Duncan et al. (1995) at 4850 MHz (PMN) with contours from 0.01 to 0.06 Jy beam$^{-1}$). A black rectangle marks the position of the slit against the distinct radio shell morphology.

**G281.9+8.7 and G283.7+9.3**. This is a faint, large, oval structure ~2.5 ° across at maximum extent discovered on SHS field HA335 (see Fig. 6) by Parker and Frew during searches for evolved planetary nebulae (e.g. Frew, Parker & Russeil 2006). The coherent H$\alpha$ morphology indicates a possible SNR candidate. Follow-up spectroscopy was undertaken at locations indicated in the Fig. 6 by white bars.

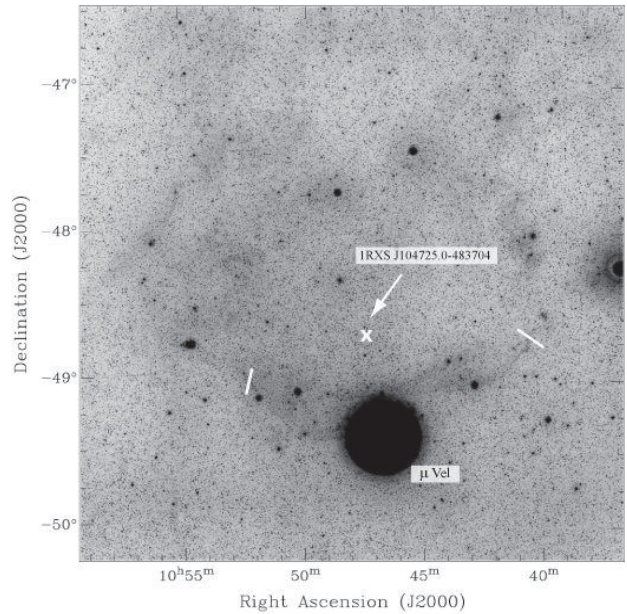

Figure 6. G281.9+8.7 and G283.7+9.3. The faint, oval structure is ~2.5 ° across at maximum extent as seen on H$\alpha$ survey field HA335. Two white bars, G281.9+8.7 and G283.7+9.3 show the position and orientation of the spectrograph slits separated by ~1.5 ° which yield a strong similarity in the observed spectra. The line ratios confirm that they are both likely part of the same object. The white cross gives the position of ROSAT X-ray source 1RXS J104725.0-483704.

Extremely high ratios of [SII]/H$\alpha$ = 3.8 and 3.4 were found in the red spectra at the 2 slit positions which are 1.5 ° apart indicating highly shocked gas, which, given the extent of the source, strongly suggests an SNR origin. Prominent [OII] at 3727Å and the presence of Balmer lines H$\beta$, H$\gamma$ and H$\delta$ in the blue were found.

SIMBAD returned a few X-ray sources within the area, but the ROSAT source 1RXS J104725.0-483704 is located directly in the center of the oval structure. Unfortunately, the available PMN and SUMSS data do not show any discernable radio structures.

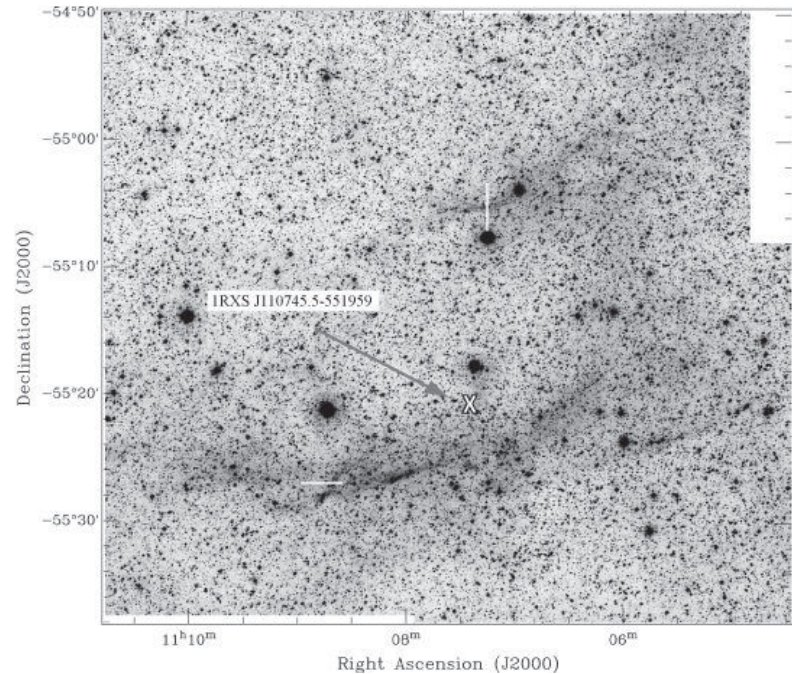

Figure 7. G288.3+4.8 and G288.7+4.5. Mosaic image (48 arcmin × 48 arcmin) of G288.3+4.8 and G288.7+4.5 built from the three high resolution images from H$\alpha$ survey field HA224. Two arcuate filaments some 5 and 15 arcmin in extent are clearly seen. Approximate slit positions are also marked. The white **X** mark indicates the position of the ROSAT X-ray source 1RXS J110745.5-551959.

**G288.3+4.8 and G288.7+4.5**

A complex of two main arcuate filaments about 15 and 5 arcmin at maximum extent were discovered on SHS field HA224. A faint background nebulosity also seems to connect these filaments (Fig. 7) on the western side and extend to the north some 7 arcmin from the smaller arcuate filament. One spectrum per filament was obtained using the MSSSO 2.3m telescope. Both spectra exhibit typical SNR characteristics with strong [SII] emission. In the blue spectra, prominent Balmer lines of H$\beta$, H$\gamma$, H$\delta$ are seen together with strong [OII] at 3727Å which also support an SNR classification. The absence of [OIII] at 4959 and 5007Å clearly rules out a PN.

Inspection of the PMN and SUMSS surveys did not reveal any significant radio detection though SIMBAD did return an X-ray source 1RXS J110745.5-551959 (see Fig. 7) located 5 arcmin north from the central region of the larger arcuate filament. A more detailed radio study at high sensitivity is required.

**G289.7+5.1**

A large, faint group of arcuate H$\alpha$ structures was discovered on SHS field HA224 (Fig. 8) where the largest filament is about 22 arcmin at maximum extent. MSSSO 2.3m blue and red spectra were obtained along the smaller arcuate 7 arcmin filament seen on the western side of Fig. 8. The flux calibrated red spectrum exhibits strong [SII] with [SII]/H$\alpha$ ~1.4 while the [OI] lines at 6300 and 6364Å are seen in the expected ratio 3:1, confirming very good sky subtraction in this decent S/N spectrum. In the blue spectrum, extremely strong H$\beta$ and [OII] at 3727Å was seen and, as is sometimes present in SNR spectra, [Ne III] +H at 3970Å.

The PMN 4850 MHz radio survey data shows a clear radio detection (Fig. 8). The radio contours (at maximum 0.06 Jy beam$^{-1}$) concentrate around and are aligned with the western 7 arcmin optical filament where the spectrum was taken and also around the longest, 22 arcmin long filament. No concentration of radio emission at 4850 MHz has been noticed in the area of the third, 15 arcmin filament, seen on the north-east side of Fig. 8.

A SIMBAD search returned the X-ray source 1RXS J111921.4-551130 located between three filaments which is marked as a `X' on Fig. 8.

**G289.2+7.1**

This is a large, prominent, arcuate filament about 30 arcmin in maximum extent and was discovered on SHS field HA278 (Fig. 9). The area some 2° around G289.2+7.1 is rich with filaments of different size and form, but this particular arcuate filament is very prominent and has a strong morphological similarity to supernova remnants. Behind the arcuate shock front, a faint trailing or ``jet'' nebulosity (Fig. 9). One red and blue spectrum, with slit position almost perpendicular to the observed shock front was obtained with the MSSSO 2.3m telescope. The red spectrum exhibits typical SNR spectral characteristics with [SII]/H$\alpha$ ~1.8 and with [OI] at 6300Å. In the blue [OII] at 3727Å is seen. Inspection of PMN and SUMSS surveys gave no detectable, coherent, radio emission.

SIMBAD returned an X-ray source 1RXS J112313.1-531345 (Fig. 9) located some 30 arcmin from the east boundary of the arcuate filament where the spectra was obtained but any association seems tenuous.

**G292.9+4.4 (RCW 59).**

This is a very large, prominent, oval nebulosity (Fig. 10) previously known as HII region RCW 59 (Rodgers, Campbell & Whiteoak 1960) located on SHS field HA225. The sensitivity of the SHS is able to reveal the full extent of this object which is now shown to be about 100 arcmin in diameter.

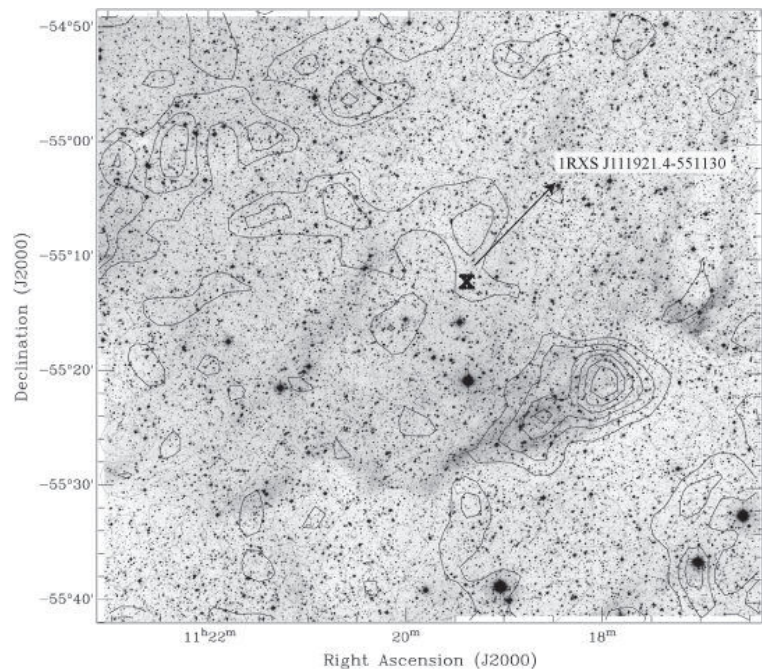

Figure 8. G289.7+5.1. A ~1° H$\alpha$ high resolution mosaic field of G289.7+5.1 with the H$\alpha$ image overlaid with PMN 4850 MHz contours in the range 0.01 to 0.06 Jy beam$^{-1}$. Three filaments 22, 15 and 7 arcminutes in extent are clearly seen. The white $\times$ sign marks the position of X-ray source 1RXS J111921.4-551130.

In the radio, SUMSS at 843 MHz covers this area, but it is not yet in the public domain. The available PMN data contoured in the range of 0.01 to 0.07 Jy beam$^{-1}$ reveals fractured structure but with an excellent match to the H$\alpha$ emission (see Fig. 10). The PMN contours continue at the same level over the south-west arcuate structure of RCW 59, where the level of H$\alpha$ emission is very low so effectively forming a complete ring of low level radio emission.

Due to the large extent of this arcuate nebulosity, SIMBAD returned more than 15 X-ray sources in the area, so it is not straightforward to deduce if any are connected with RCW 59. One pulsar, J1132-5627, is situated in the north-west arc of RCW 59.

The new MSSSO 2.3m blue and red spectra exhibit typical SNR characteristics with [SII]/H$\alpha$ ~0.97 and weak [OIII] at 4959 and 5007Å relative to H$\beta$ (which rules out an evolved PN), H$\gamma$ and H$\delta$, and very strong [OII] at 3727Å. The NII/H$\alpha$ ratio is ~0.7 at the very upper limit of what is found in HII regions (Fesen, Blair & Kirshner 1985). Also in the blue spectrum lines of Ne+H at 3888Å and [NeIII]+H at 3970Å are detected which are sometimes present in SNR spectra.

This object was known as a star forming HII region or possibly part of some larger structure (Georgelin et al. 2000; Avedisova & Palous 1989), most likely due to morphological similarity of the RCW 59 optical loop with other HII regions. No optical spectra could be found in the literature, so ours is the only existing spectrum. Strong support for re-classification of RCW 59 as a SNR comes not only from this new spectrum and the associated radio structure seen with the PMN but the morphology of the more extensive optical nebulosity now revealed by the SHS. Additional spectra along more of the arcuate structure and especially in the south-east part of RCW 59 (diffuse emission area) are required for a final conclusion about the nature of this object.

The Galactic coordinates presented are for the central area of RCW 59 (from Rodgers, Campbell & Whiteoak 1960) and not coordinates of the slit position as generally given in this Catalogue.

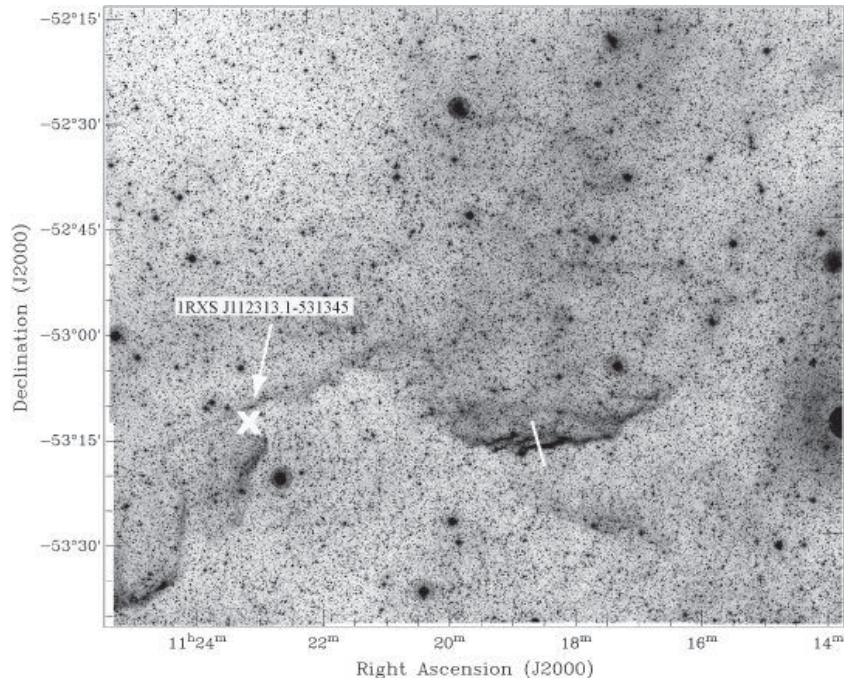

Figure 9. G289.2+7.1. High resolution mosaic image of G289.2+7.1 built from 12, 30×30 arcmin high resolution SHS images. The slit position of G289.2+7.1 was almost perpendicular to the 30 arcmin filament front shock. Faint ``trailing jets'' can be seen behind this front shock which is common in some remnants. Again the × sign marks the position of the ROSAT X-ray source 1RXS J112313.1-531345.

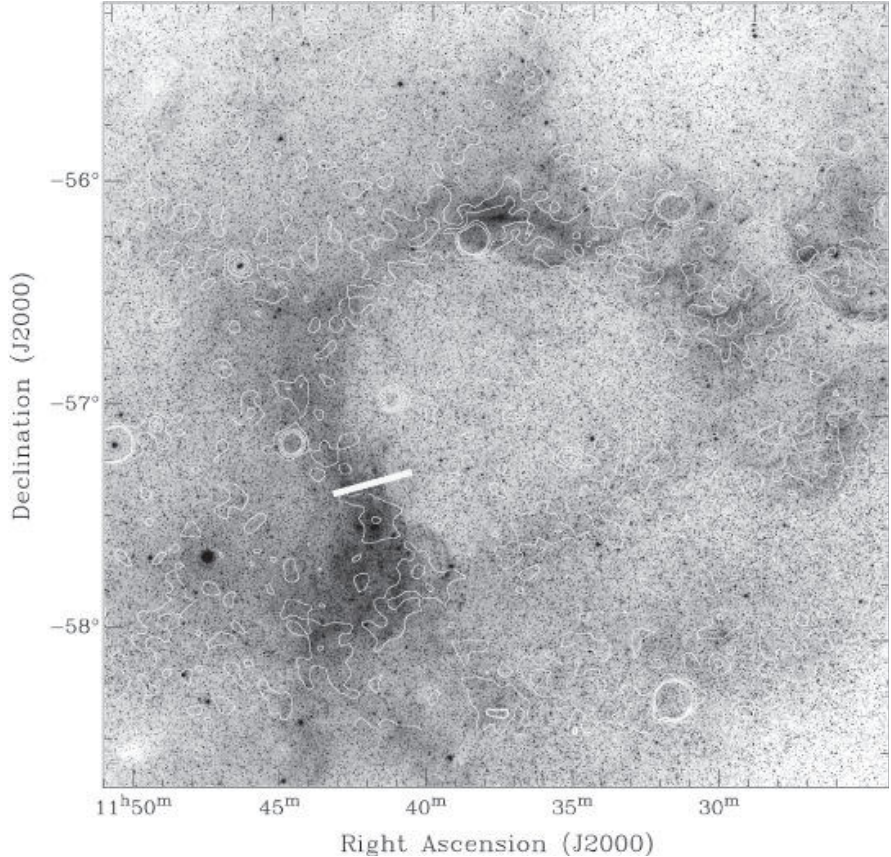

Figure 10. The H$\alpha$ emission of G292.9+4.4 (RCW 59) from the H$\alpha$ survey overlaid with radio contours (0.01 to 0.07 Jy beam$^{-1}$) from the PMN survey (4850 MHz). It is clear that the radio data, though fractured, closely follows the arcuate structure seen in H$\alpha$.

### G303.6+5.3

This is a large, prominent, oval formation about 4° ×2.5° in extent discovered on SHS field HA228 (Fig. 11). SIMBAD returned two X-ray sources (ROSAT 1RX J131108.3-575012 and 1RX J130518.6-575816) in the central area separated by about 45 arcmin. Towards the western edge there is a pulsar J1317-5759 (Fig. 11). These two X-ray sources and pulsar are strongly indicative of a supernova remnant in this region, but further analysis is required to reveal any true connection. More tellingly, PMN 4850 MHz radio data (see Fig. 11), though fractured, closely follows the oval structure seen in the H$\alpha$ survey. The SUMSS 843 MHz survey data of this region are not yet in the public domain to offer corroboration and an opportunity to measure a spectral index.

MSSSO 2.3m blue and red spectra have been obtained (see slit position at Fig. 11) which exhibit typical SNR characteristics. In the red spectrum strong H$\alpha$ and [SII] are present with [SII]/H$\alpha$ ~ 0.74. The blue spectrum has strong [OII] 3727Å and H$\beta$, H$\gamma$ and H$\delta$ Balmer lines. Also, there is weak but clearly present [Ne iii] +H at 3970Å. All these lines sustain the SNR classification.

### G306.7+0.5

This is a large, prominent, arcuate nebulosity about 25 arcmin N-S at maximum (Fig. 12) discovered on SHS field HA180. The Wolf-Rayet star (van der Hucht 2001) WR 53 (HD 117297) is about 50 arcmin eastward. At 40 arcmin in the same direction the ROSAT X-ray source 1RXJ J132850-615458 is located and about 3 arcmin further south-east is the pulsar J1329-6158 (see Fig. 13).

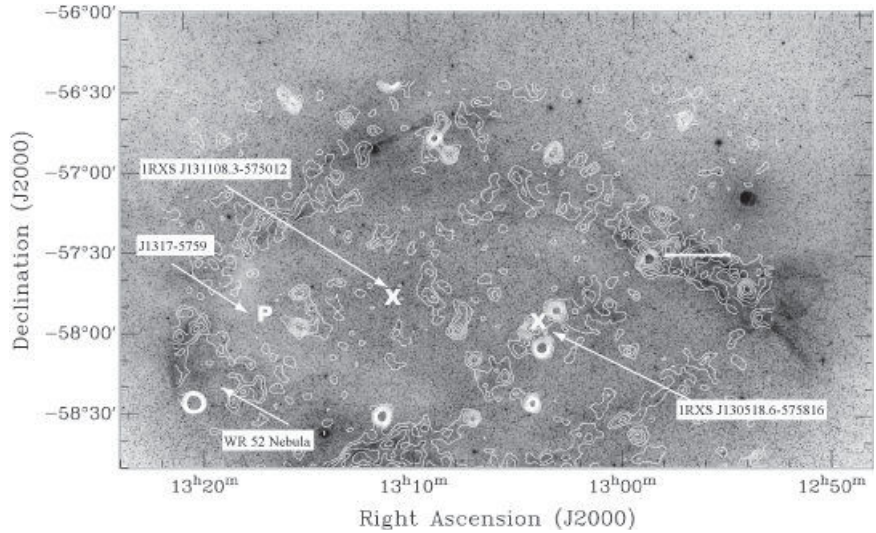

Figure 11. G303.6+5.3. The large H$\alpha$ oval formation ~4°×2.5° in size, with all related objects indicated. The white × mark the positions of two ROSAT X-ray sources while the **P** indicates the pulsar J1317-5759. On the southeast side is marked the Wolf-Rayet star nebula WR 52, evidently not connected with this SNR candidate. The slit position is marked on the west side of the image. The optical oval structure is overlaid with PMN 4850 MHz contours from 0.01 to 0.07 Jy beam$^{-1}$, which, although fractured, closely follows the H$\alpha$ emission.

Although the area is complex and hard to disentangle the existence of the X-ray source and pulsar and the arcuate structure to the west indicated that the nebula might be part of an SNR candidate worthy of spectroscopic follow-up. Blue and red 2.3m spectra were obtained. The flux calibrated red spectrum gave [SII]/H$\alpha$ = 0.9 which is more the signature of an SNR but such large values are not unknown in Wolf-Rayet ejecta. The spectrum also exhibited [OI] at 6300 and 6364Å in the expected lock-step ratio of 3:1 while in the blue the presence of all prominent Balmer lines and [OII] at 3727Å also support an SNR classification.

Inspection of the PMN 4850 MHz and SUMSS at 843 MHz radio survey data shows some very weak radio detection, (almost at the level of the noise), so precise details cannot be deduced about the veracity of the radio detections until more sensitive observations are carried out.

**G315.1+2.7.**

This is an isolated, filamentary nebulosity, extending some 11 arcmin in the north-south direction and first seen as an elongated nebular ``blob'' in the 16× blocked-down fits image of SHS field 231. Examination of the original survey films at ROE showed the fine, twisted structure and full extent of this filament presented at full SHS resolution in Fig. 14.

Investigation showed that this filament is the optical counterpart of a possible SNR candidate G315.1+2.7 identified from previous radio observations (Duncan et al. 1995, 1997). An excellent match was achieved between the optical filament and radio flux contours on the east part of G315.1+2.7, especially with the radio map at 843 MHz (SUMSS) (see Fig. 14).

Optical spectra, taken at two locations along the filament had typical SNR characteristics such as strong [SII] but with different values of the [SII]/H$\alpha$ ratio of 0.8 and 1.5. In the blue, both spectra showed all Balmer lines and [OII] at 3727Å. This object, which we now confirm is a bona-fide Galactic SNR, forms the basis of a detailed analysis in Stupar, Parker & Filipović 2007a.

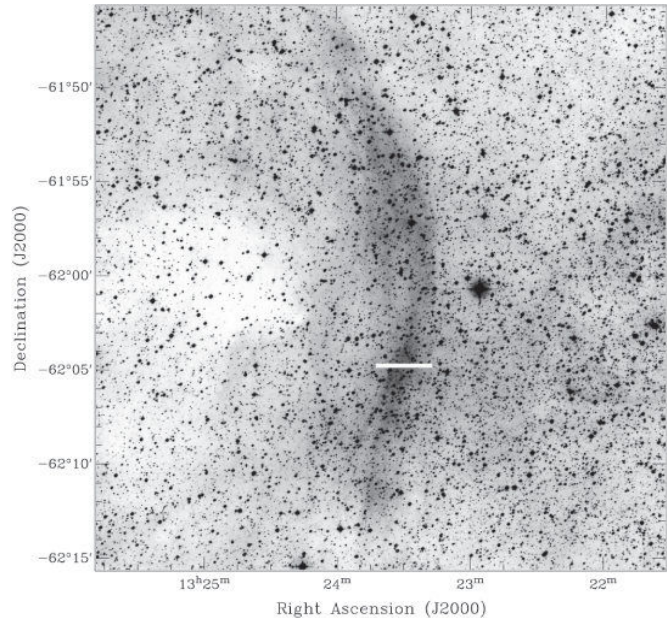

Figure 12. G306.7+0.5. A 30 arcmin SHS H$\alpha$ image of the large arcuate emission structure. The slit position is indicated with a white bar and was oriented E-W at 90 degrees to the filament.

**G332.4+0.6**

This is a large, almost complete ring of H$\alpha$ emission, approximately 30 arcmin in diameter but with a strong, complicating HII region on the north-east side (Fig. 16). It was uncovered on SHS field HA289. The oval form indicated it might be an SNR and a 2.3m optical spectrum was obtained.

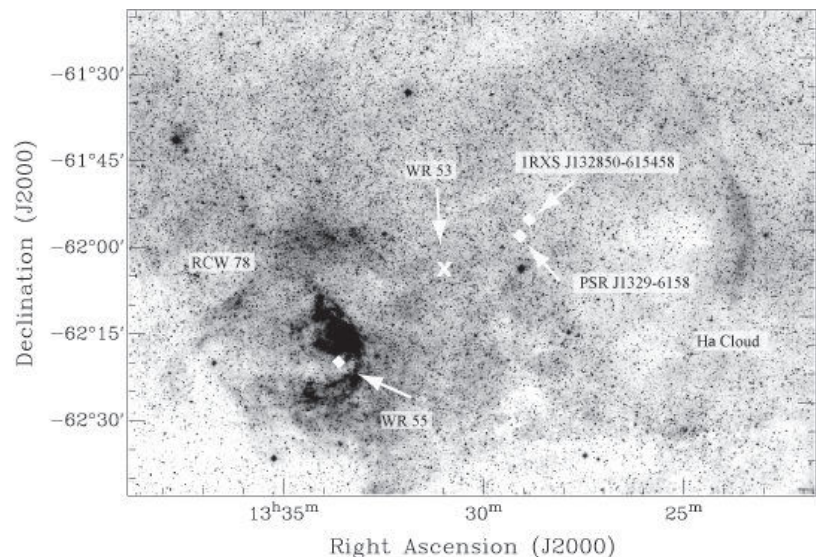

Figure 13. A much larger H$\alpha$ image of the area around G306.7+0.5 showing the position of the X-ray source 1RXS J132850-615458 and pulsar PSR J1329-6158. The position of the Wof-Rayet star WR 53 is also marked as a possible source of the observed H$\alpha$ arcuate nebulosity. The image also shows the position of WR 55 and surrounding nebula RCW 78.

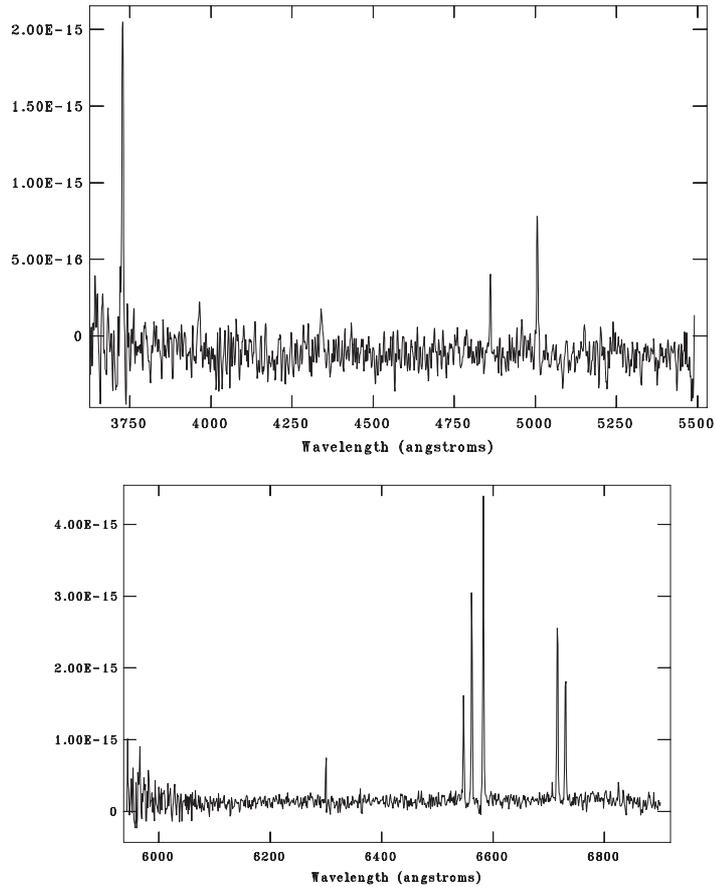

Figure 15. Blue and red flux calibrated spectra for G315+2.7. revealing in the red strong [SII] lines, the strong [NII] relative to H$\alpha$, and the [OI] lines at 6300 Å. The blue spectrum (top image) has the [OIII] doublet at 4959 and 5007Å showing ~200 against H$\beta$ =100.

      The red spectrum exhibits typical SNR spectral characteristics placing it firmly in the SNR domain. The strong [NII] / H$\alpha$ ratio is also not seen in HII regions while in the blue, the spectrum has low S/N but the only easily recognized line is H$\beta$ so the lack of detectable [OIII] also mitigates against the object being an evolved PN.

      A SIMBAD search returned a pulsar J1610-5006 located on the northern limb of this ring nebula. In the radio, the PMN 4850 MHz survey data reveals a striking ring structure which closely follows the H$\alpha$ emission (Fig. 16) though this was not previously identified as an SNR candidate. The SE part of this structure is without PMN data so confirmation that the radio data completely follows the H$\alpha$ emission is not currently possible. Nevertheless the strong co-incidence between the shell-like optical and radio morphology coupled with the highly indicative optical spectroscopy enables identification as a new Galactic SNR.

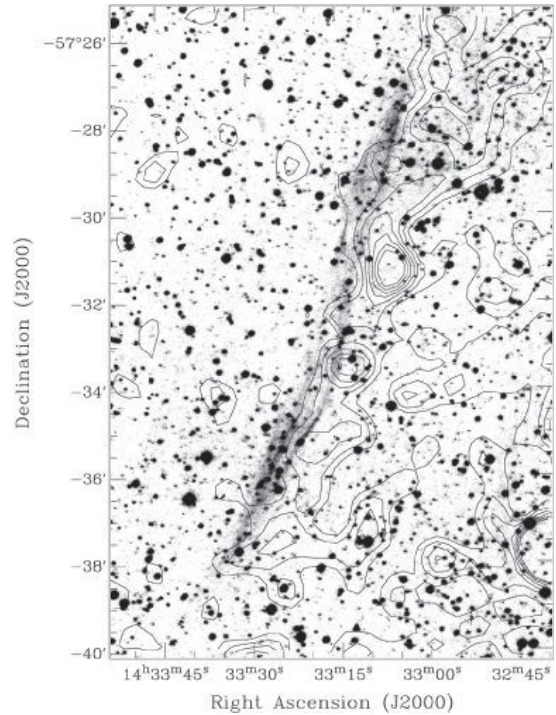

Figure 14. G315.1+2.7 filament as a quotient image ( H$\alpha$ divided by the matching short red) overlaid with SUMSS 843 MHz contours from 0.001 to 0.01 Jy beam$^{-1}$. There is clear alignment between the radio and optical emission with a displacement between the radio and optical filament of < 1 arcmin.

**G343.4+3.6.**

This object comprises two broad, arcuate filaments (Fig. 17) about 7 and 4 arcmin at maximum extent which were discovered on SHS field HA428. This field contains a significant amount of complex H$\alpha$ emission structures but these stood out as a decent SNR candidate. Low resolution SAAO 1.9m spectra (300 lines/mm covering 3500-7500Å) were taken across these newly identified filaments.

The spectra do exhibit SNR characteristics. All prominent Balmer lines are registered as well as [OII] at 3727Å. The lack of [OIII] in the blue mitigates strongly against a PN identification and though [SII]/H$\alpha$ = 0.6 is at the lower end of values found for SNRs the [NII]/H$\alpha$ ratio of 0.8 is greater than that found in HII regions. The spectra were taken on a non-photometric night so the extinction estimates are indicative (extinction coefficient $c$ =1.26) though the H$\alpha$ to H$\beta$ ratio is » 3 so modest extinction is present.

SIMBAD did not return any interesting objects in the vicinity but inspection of the PMN data does detect the object. Fig. 17 shows the radio contours of 0.01 to 0.07 Jy beam$^{-1}$ which closely matches the underlying H$\alpha$ emission making an association highly likely. More precise details are not possible due to the low resolution of the PMN survey (FWHM of 64 m dish was ~4.3 arcmin) but the combination of optical spectral properties, elimination of most likely alternatives and matching radio and optical emission permits identification as a probable new Galactic SNR.

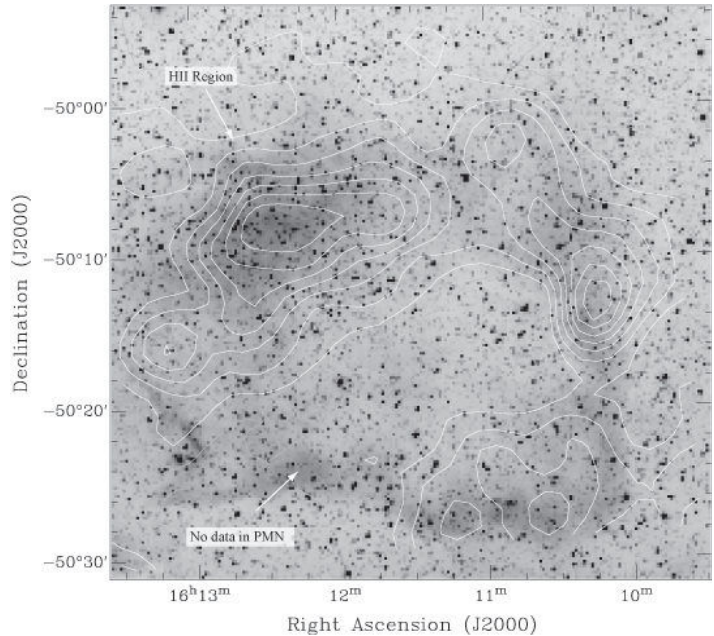

Figure 16. G332.4+0.6 in H$\alpha$ light with overlaying PMN contours (from 0.01 to 0.14 Jy beam$^{-1}$). The unrelated HII region is indicated to the immediate N-E. In the SE the PMN survey is without data so it is not yet possible to confirm a complete match of the radio and optical emission across the entire ring.

### G332.5-5.6 - The ``Paperclip''.

This SNR was one of the first to be optically detected from the H$\alpha$ survey during a search for PN candidates as part of the MASH project. This remnant was first recognized as a candidate SNR in Duncan et al. (995) due to its approximately 30 arcmin circular radio morphology. In Parker, Frew & Stupar (2004) we presented the first optical spectra of one of the newly identified filaments of G332.5-5.6 which indicated a likely SNR origin. Full details can be found in Stupar et al. (2007c) where the non-thermal nature of this object was confirmed. For convenience and completeness here we present in Fig. 18 an H$\alpha$ image of the brightest optical filament where the nebulosity has the form of a ``paperclip''. This object was independently studied by Reynoso & Green (2007) but based solely on radio data though this object was recognised first by our SHS imagery.

### G329.9-7.8

This is a prominent, arcuate nebulosity about 7 arcmin at maximum extent discovered on SHS field HA236. Figure 19 shows two trailing amorphous ``jets'' some 10 arcmin behind the main shock front in the south-easterly direction. Typically, these trails, seen in many SNRs, are located right behind the shock front while here they are ~10 arcmin behind so a definite connection is not proven. The H$\alpha$ structure indicated it might be an SNR and worthy of spectroscopic follow-up.

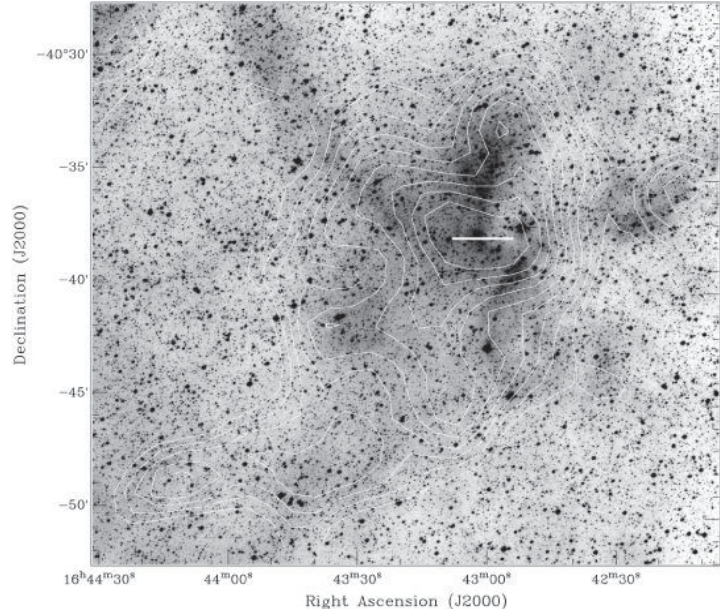

Figure 17. G343.4+3.6 with slit position and PMN contours (0.01 to 0.07 Jy beam$^{-1}$) overlaid on the H$\alpha$ image. Strong correlation between the optical and radio emission is evident.

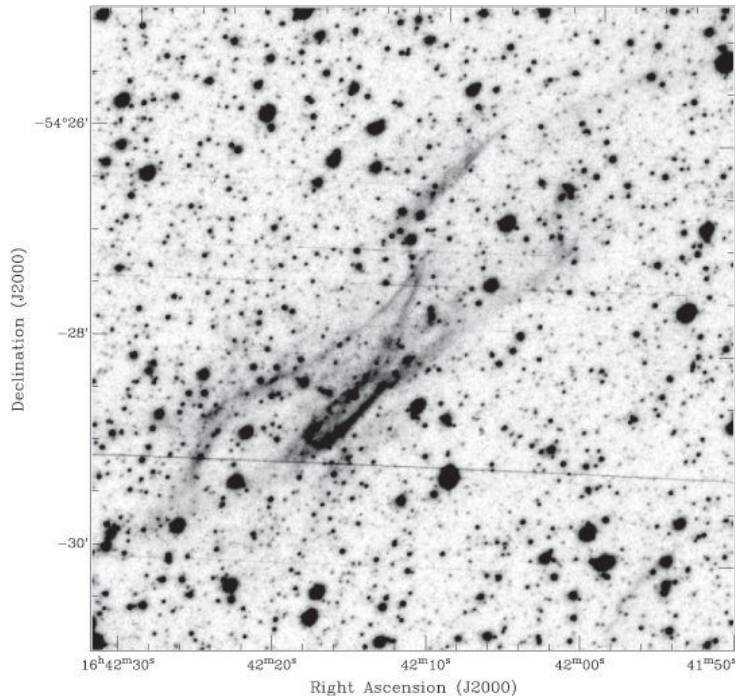

Figure 18. The most prominent optical filament of G332.5-5.6 in the form of a ``paperclip''. For full details see Stupar et al. (2007c).

An SAAO 1.9m low resolution spectra was obtained and the slit position marked on Fig. 19. The spectrum exhibits typical SNR characteristics with [SII]/H$\alpha$ ~ 1.18 and [NII]/H$\alpha$ ~ 1.8, effectively eliminating an HII region and PN. [OI] at 6300Å was also present while in the blue strong [OII] at 3727Å is detected, important in aiding SNR classification. The spectrum was taken on a non-photometric night of poor seeing so a formal extinction calculation was not performed though the weakness of H$\beta$ indicates moderate extinction.

A SIMBAD search did not return any interesting object in this area and nothing was noted in the existing radio data.

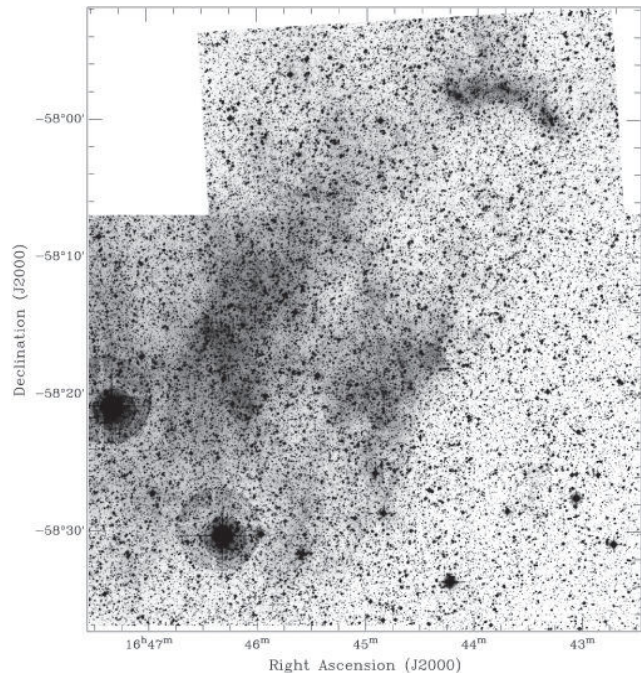

Figure 19. G329.9-7.8. This is the prominent, arcuate nebulosity at the top right hand corner of the large 45 arcmin mosaic image. The lower left nebular clouds may represent ``jet trails'' from the possible SNR shock front.

**G348.2-1.8**

This unusual object, discovered on SHS field HA483, is represented by two prominent, parallel arcuate nebulosities (Fig. 20) about 8 and 5 arcmin at maximum extent and separated by about 2 arcmin. One 2.3m spectrum was obtained for each filament which exhibit typical SNR characteristics with [SII]/H$\alpha$ ~ 0.7-0.9 together with H$\beta$, and [OII] at 3727Å in the blue part of the spectrum. The [NII]/H$\alpha$ ratio from both slit positions varies from 0.8 to 1.1 effectively ruling out a HII region origin.

SIMBAD did not return any remarkable object in this area but the PMN 4850 MHz radio survey data shows an object co-incident with the H$\alpha$ emission with the same orientation and of slighter larger extent at about 13 arcmin. A wider 25×32 arcmin region is shown in Fig. 21 which reveals a more extensive radio ring indicating that a much larger candidate may be present. This area is not yet scanned by SUMSS while the VLA NVSS 1.4 GHz survey does not show any particular object. More sensitive radio maps of this region are required.

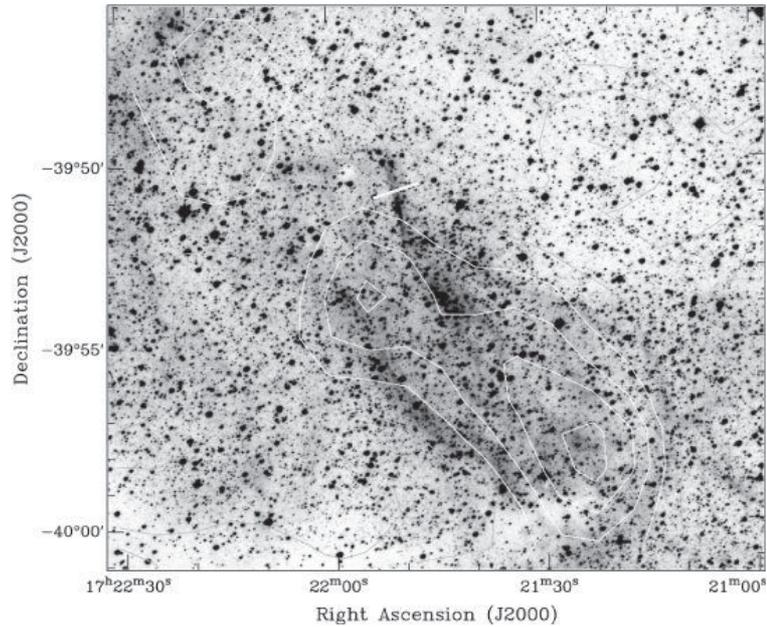

Figure 20. G348.2-1.8, exhibiting two arcuate nebulosities with slit position for the northernmost component indicated by the white bar. The sizes of these parallel nebulosities are 8 and 5 arcmin at maximum extent and are separated by about 2 arcmin.

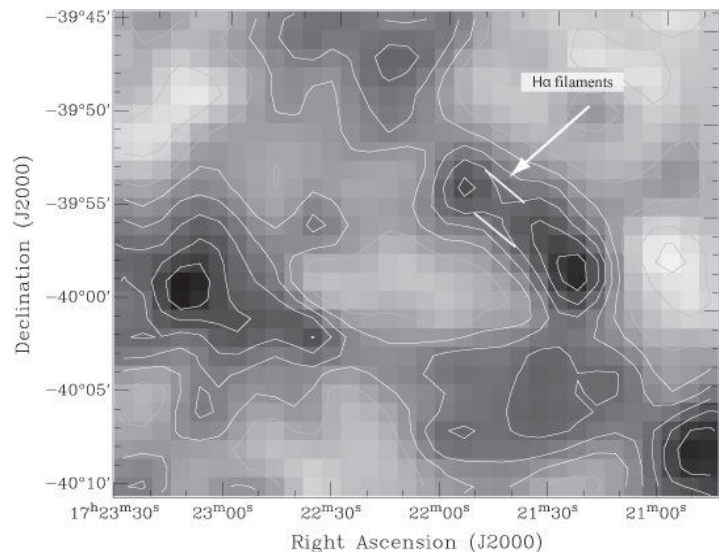

Figure 21. Wider area around G348.2-1.8 with the position of the H$\alpha$ filaments indicated. Without higher sensitivity and improved resolution radio observations, firm conclusions about the connection of the extended radio sources into a coherent shell-like remnant cannot be made but the PMN data is highly suggestive. The PMN data contours are from 0.01 to 0.05 Jy beam$^{-1}$.

**G18.7-2.2**

This object was first identified as an extremely large, arcuate filamentary nebulosity about 30 arcmin at maximum extent (Fig. 23) on SHS field HA1060. However, a high resolution SHS mosaic shown in the left panel of Fig. 23 shows a much wider area, some 2° in extent in declination revealing several large, nebulous regions. To the west is known Galactic SNR G18.9-1.1 which shows up as a compact and intense region of radio emission in the PMN 4850 MHz radio map in the right panel of Fig. 23. This remnant is probably unrelated to the other more extensive, fainter and surrounding radio and nebular structures which show a high degree of positional coincidence and which we consider to be another, older remnant. There are two parallel radio and optical structures about 45 arcmin in extent. Off the south east edge of the more southerly structure at nearly right angles is the originally noticed north-south filament. The radio structure along this filament is more fractured. The NVSS survey shows no data in this area. It is not possible to yet make a definitive connection between the detached but parallel northernmost nebulous cloud and the connected Southern elongated components which lie between -12°36′ and -13°57′. This is where the optical spectra were obtained.

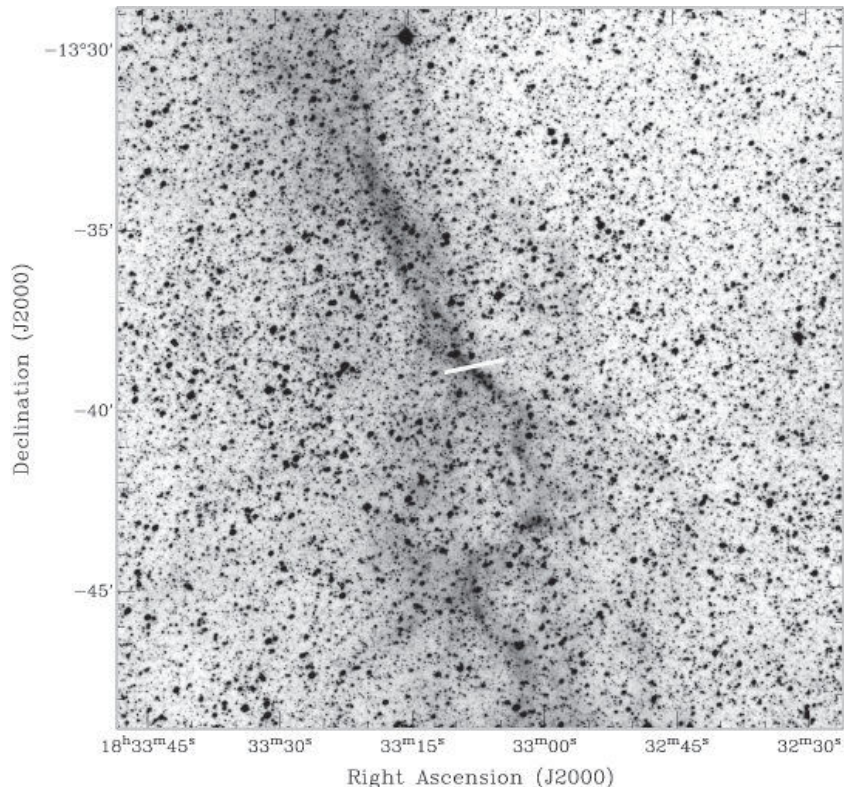

Figure 22. G18.7-2.2. South part the filamentary nebulosity with slit position indicated. It is possible that this is only part of a larger arcuate nebulosity of 1.5° in extent. More spectra are required for a final conclusion.

Red and the blue MSSSO 2.3m spectra were obtained which exhibit typical SNR characteristics with [SII]/H$\alpha$ ~ 1.33. In the blue, the strongest lines are H$\beta$ and [OII] at 3727Å but [OIII] at 5007Å, H$\gamma$, H$\delta$, He+H at 3888Å and [NeIII] +H at 3970Å are also present. The slit position is marked on Fig. 22. The [NII]/H$\alpha$ ratio of ~0.72 is at the extreme of what has been found in HII regions and the morphology alone rules out a possible evolved PN. These preliminary optical spectra for one small region of this complex is highly suggestive of an SNR but additional spectra across the filaments are needed for confirmation. An estimate of radio spectral index to reveal the non-thermal nature of the structures would be valuable.

SIMBAD returned some X-ray and PMN point-like sources, as well as pulsars in the area, but due to the large size of this nebulosity it is not yet possible to confidently assign any of these sources to this large remnant.

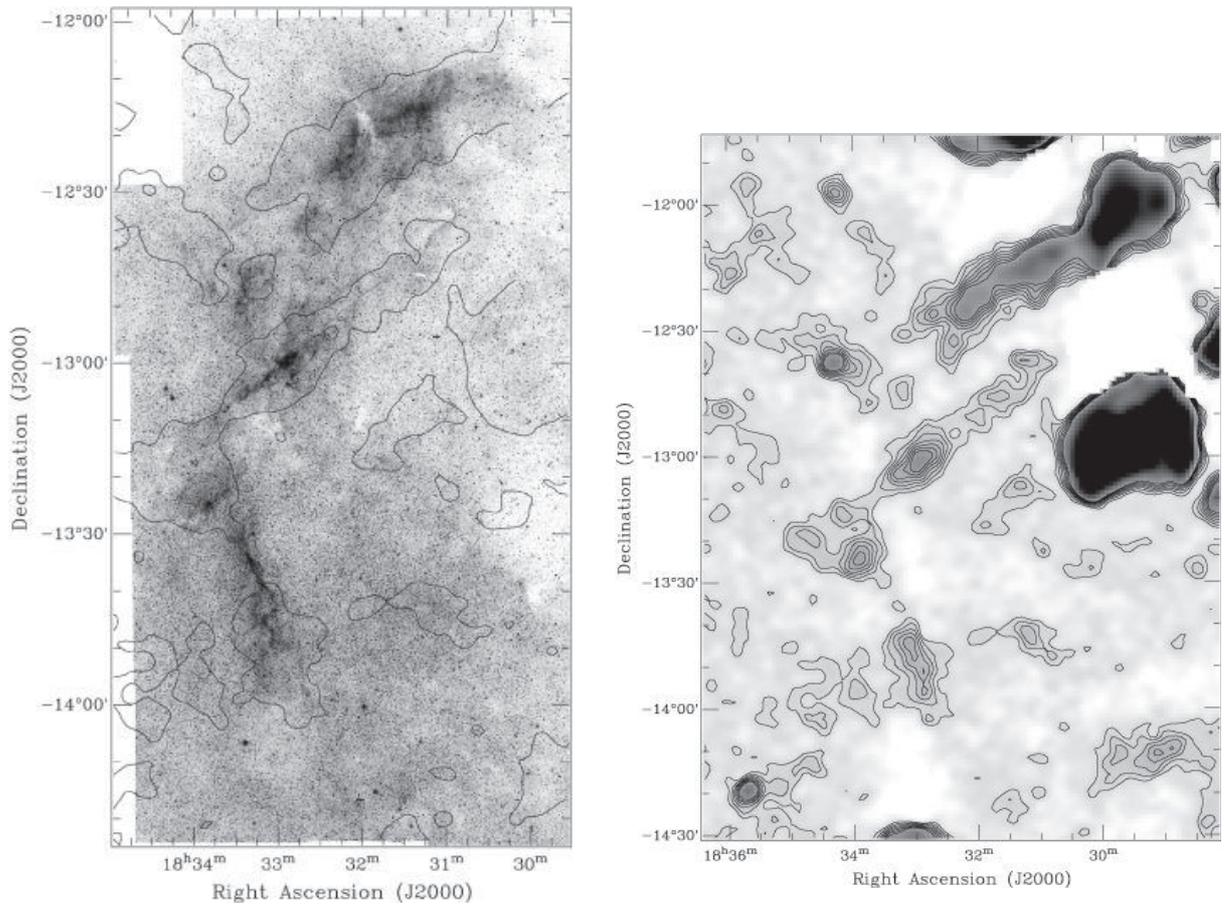

Figure 23. The left image, 2° in extent, is the high resolution SHS mosaic around slit position G18.7-2.2 2. The lowest level radio contours of 0.01 Jy beam$^{-1}$ are taken from the PMN 4850 MHz radio survey. These highlight the excellent match of the radio and optical emission. The right image is a PMN radio image of the same area also overlaid with radio contours from 0.01 to 0.08 Jy beam$^{-1}$. The southern arcuate nebulosity consists of many PMN point-like sources. The available optical spectra confirm shock excited gas typical for SNRs. The strongest, compact source to the west is known Galactic SNR G18.9-1.1.

## 4 Discussion

### 4.1 Morphological structure and Galactic distribution of the new remnants

From the 18 new Galactic SNR candidates uncovered in this work from the SHS and for which strongly indicative spectroscopic confirmation has been obtained, 10 of them have H$\alpha$ emission in the form of a shell (or extended arc; see Table 2). This provides strong, independent morphological support towards their identification as SNRs (Lozinskaya 1992). About 50% of the new remnants also have corroborating radio detections at single and in some case multiple frequencies (example G303.6+5.3, G348.2-1.8; see the Catalogue). These radio detections more or less match the optical emission in terms of alignment and extent. For some of them (eg. G332.5-5.6) non-thermal radio spectral indices are also confirmed. This group of observed, shell-like, optical remnants show breaks in the shell structure due to thermal instability and inhomogeneities of the interstellar medium which is typical for SNRs. Behind the evident shock fronts, some of the remnants (e.g. G243.9+9.8 or G289.2+7.1) appear to possess trailing jets of emission which are also present in some known SNRs. An example of the excellent match in morphological structure of these new shell-like remnants is between new candidate SNR G243.9+9.8 and well known Galactic SNR IC 443.

It is the SHS imagery which has now enabled an association to be made between new, optically uncovered candidate SNRs and the often weak and fractured radio detections which had previously not been recognised or considered as part of any possible SNR.

Table 2. Morphological structure of the observed filaments

| SNR candidates in the form of shell (or arc) | SNR candidates with irregular form |
|---|---|
| G253.0+2.6 | G283.7-3.8 |
| G243.9+9.8 | G288.7-6.3[a] |
| G281.9+8.7 | G289.7+5.1 |
| G288.3+4.8 | G343.4+3.6 |
| G289.2+7.1 | G332.5-5.6[a] |
| G292.9+4.4 | G351.1+4.9 |
| G303.6+5.3 | G348.2-1.8 |
| G306.7+0.5 | G18.7-2.2[b] |
| G332.4+0.6 | |
| G329.9-7.8 | |

[a] SNR (or candidate) shell in the radio but irregular in the optical.
[b] Due to the large angular size, classification is difficult but it could also be considered as part of a very large shell structure.

The remaining 10 out of 18 SNR candidates do not constitute recognisable morphological structures such as a shell or coherent arc but are just localised emission nebulosities or apparently isolated H$\alpha$ filaments of different sizes. In many cases the full extent of any optical counterpart to these remnants may be hidden by intervening dust or is not present due to the variable density of the surrounding ISM. Nevertheless, their optical spectra exhibit typical SNR emission lines while other observed line ratios usually rule out HII regions or PNe. Other possible emission nebulae such as Wolf-Rayet shells or Herbig-Haro objects are also easily discounted based on environment, morphology and lack of an obvious ionising point source. This was established after checking known hot star catalogues or from examination of the broad-band colours of stars in the immediate vicinity of the nebulae using the SuperCOSMOS sky survey data (e.g. Hambly et al. 2001) to look for very blue central objects.

Even in cases where the optical emission is patchy or fragmented it still matches equivalent radio emissions when detected (eg. see G289.7+5.1 and G315.1+2.7). SNR G332.5-5.6 is an excellent example of such fragmented optical filaments which still produce a good match to the equivalent radio and X-ray data. Together with the optical spectra, a compelling case for SNR identification is made (see also Stupar et al. 2007c).

For some of the newly detected optical filaments, especially those without radio detection (or an X-ray source within the optical emission zone), it is very hard to make any firm conclusion on the true origin of the emission even if their spectra show typical SNR characteristics. There are two possible solutions to what they might represent.

First, they could be the very last vestiges of very old SNRs is in the final, dissipation phase where they are almost completely mixed with the ISM. This is all that can be seen for some senile SNRs in the optical (mostly H$\alpha$). No other kind of radiation can be expected (and particularly not in the radio; see later discussion). Second, they may form part of a remnant in the snowplow phase and, depending on the nature of interstellar medium, (mostly density and temperature) H$\alpha$ radiation can arise. In this case, only fragmented optical filaments in H$\alpha$ may be seen but in the radio (or in X-rays) the whole or partial structure could be expected to be detected (an example is G332.5-5.6).

The diffuse, shell-like, morphological structures, short filaments or small, localised nebulosities, can in principle, be present in young as well as in old SNRs. However, based on the overall structure of all 18 observed objects, most are clearly old. Young SNRs, like Cassiopeia A, show very compact and bright (but very complex) shells where inhomogeneities of the interstellar medium cannot have a strong influence in such a very ``short'' lifetime of a few hundred years. Of course, this is not a strict rule as observed angular ``compactness'' depends on distance as well as age and, to a lesser extent, also on local, small-scale ISM inhomogeneities.

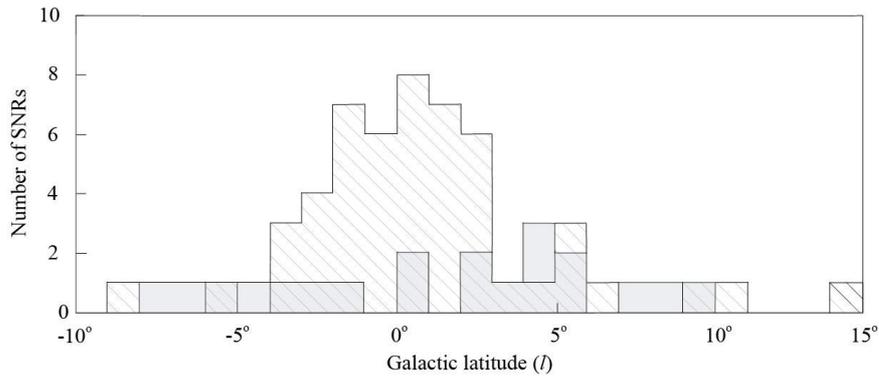

Figure 24. The number of optically detected Galactic SNRs as a function of Galactic latitude (*l*). The diagonally filled histogram is for the 52 previously known SNRs optically detected and catalogued in Green (2006). The shaded histogram is for the 18 new optically detected SNR candidates discovered in this work. Green's histogram shows a concentration of optically detected SNRs around the Galactic plane at low latitudes. There is one previously known optically detected remnant at $l \sim 15°$, SNR G327.6+14.6, (the remnant of supernova 1006) which, as a young remnant, also has published optical observations.

Many of these new, optically detected SNR candidates are at relatively high Galactic latitudes $|b|$. Only 3 from 18 candidates have $|b| < 1°$ but essentially the distribution is flat across the SHS full Galactic latitude range out to $|b| \sim 11°$ (see Fig. 28). It is a little suprising more new SNR candidates have not been found closer to the Galactic plane at low latitudes where the concentration of previously known optically detected remnants is higher. Strong extinction clearly plays a role. According to Green (2006), the ratio between objects close to the Galactic plane ($b \pm 1°$) and those with $b > \pm 1°$ is approximately 2.5:1 in favour of remnants concentrated near the Galactic plane.

Another problem for objects at higher Galactic latitude is the contradiction with kaf80 and the fact that the shock must collide with a dense (~0.15 electrons cm$^{-3}$) ISM in order for optical emission lines to be detected. There is a concentration of dense clouds around the Galactic plane, with a decrease of ambient density on both sides of the plane with increasing Galactic latitude. Furthermore, Balmer lines, seen in the optical spectra of the entire sample, rise in strength in a high-density environment. At higher latitudes, the ambient density is much lower so the local medium around the newly detected optical remnants must represent an above ambient density responsible for collisional excitation, creation of a shock wave and the Balmer line visibility. Another possibility is that most of these new SNRs are local and so can appear at higher latitudes and still remain within the dust and gas layer. The high sensitivity of the SHS is simply enabling detection of such local, more evolved remnants than before. Independent distance estimates for these remnants is urgently required. The putative $\Sigma - D$ relation, (Stupar et al. 2007d), does not have the requisite accuracy to clearly solve the ambiguity.

### 4.2 Emission lines analysis

The detected remnant optical emission lines provide vital clues to their physical nature and a powerful diagnostic capability, eg., phillips99, ken00 and riesgo02. Alongside the strength of the [SII] lines relative to H$\alpha$ the key indicator of SNR shocked material, [OI] and [OII] and [OIII] lines are important. The [OII] 3727Å doublet is seen in practically all SNR spectra from the sample and is usually very strong in the flux calibrated spectra despite both the poor CCD blue performance and variable extinction giving low S/N shortward of 4000Å. Only G329.9-7.8 is without obvious [OII] at 3727Å. This non detection is not definitive since the observations were done under non-photometric conditions and extinction here plays a significant role. When resolved the [OII] doublet at 3727 and 3729Å can be used to estimate electron density $n_e$ (Raymond 1979; Osterbrock & Ferland 2006). The spectral resolution employed in this work was insufficient to separate these two lines but the [SII] doublet, which are well resolved and of good S/N can also be used for density estimates (Osterbrock & Ferland 2006).

The oxygen [OI] lines at 6300 and 6364Å are usually present in the sample which lends additional support to an SNR classification. However, these lines are also strong in the night-sky spectrum. Hence, given the relatively low spectral resolution and the often highly extended nature of the nebula along the slit (limiting scope for extracting a suitable sky region of sufficient S/N), these lines often suffer imperfect sky-subtraction with the expected 6300/6364Å 3:1 ratio not often observed. The objects are also Galactic so there is no significant velocity difference to offset these lines from the night sky except at higher dispersion. Due to these difficulties these lines were not used in any subsequent analysis. Nevertheless, their obvious detection is a strong supporter of an SNR origin as these lines are rarely seen in HII regions and weak/absent in PNe, the most likely alternative possible identifications. The [OIII]lines at 4959 and 5007Å in combination with [OIII]at 4363Å can provide nebula (electron) temperature estimates with a ratio of 4959 + 5007Å/4363Å $<$ 30 is regarded as an excellent indicator of an SNR as such a high ratio, indicative of high electron temperatures, cannot be observed in HII regions or usually PNe (Fesen, Blair & Kirshner 1985). However, only 12 objects from the observed candidate sample have detectable [OIII] 4959 and 5007Å lines with the [OIII] line 4363Å being far too weak to be detected.

In most cases the 4959 and 5007Å lines are observed in the expected lock-step ratio of 1:3 but in the case of G281.9+8.7 only a weak 5007Å component is detected due to low S/N. The general weakness of [OIII]emission in the candidate remnants compared to strong [OII] 3727Å rules out the nebula being evolved PNe unless of very low excitation. Furthermore, although the [OIII]line at 5007Å is frequently detected in HII regions, where they are usually weak compared to H$\beta$ the observed [NII]/H$\alpha$ ratio is usually much greater or at the limit of that for HII regions (e.g. Kennicutt 2000) except for G288.7-6.3 (0.46), G306.7+0.5 (0.55) and G303.6+5.3 (0.58). In the latter 2 cases the [SII]/H$\alpha$ ratio is also much greater than encountered in HII regions or PN.

Detection of [OIII] 5007Å is common in SNRs and can be produced in shocks with velocities over 70-80 kms$^{-1}$ (Raymond 1979). This occusr because slower shocks cannot produce doubly ionized oxygen, while those faster than 140 kms$^{-1}$ produce so much H$\beta$ that the oxygen line becomes relatively weak (Raymond 1979). The [OIII] lines observed in SNRs can give us some idea about shock velocities and, when other data and line ratios have eliminated HII regions or PNe, they remain consistent with an SNR identification.

Table 3 Selected physical estimates for the observed sample for every slit position

| Name | Shock velocity $V_s$ (kms$^{-1}$) | Electron density $n_e$ (cm$^{-3}$) |
|---|---|---|
| G253.0+2.6 | 88 | 181 |
|  | 100 | 626 |
| G243.9+9.8 | 70 | - |
| G283.7-3.8 | 85 | 33 |
|  | - | 129 |
|  | 82 | 139 |
| G288.7-6.3 | 82 | 760 |
| G281.9+8.7 | 80 | 170 |
| G288.3+4.8 | 50 | 3.64 |
| G289.7+5.1 | 50 | 41 |
| G289.2+7.1 | 100 | 139 |
| G292.9+4.4 | 85 | 41 |
| G303.6+5.3 | 50 | 26 |
| G306.7+0.5 | 50 | 74 |
| G315.1+2.7 | 95 | 33 |
|  | >40 | - |
| G332.4+0.6 | 70 | 240 |
| G343.4+3.6 | 50 | 83 |
|  | 110 | 336 |
| G332.5-5.6 | 82 | 227 |
|  | 98 | 119 |
|  | 84 | 49 |
|  | - | 400 |
|  | - | 600 |
|  | - | 49 |
| G329.9-7.8 | 80 | 159 |
| G348.2-1.8 | 60 | 57 |
|  | 40 | - |
| G18.7-2.2 | 50 | 92 |

### 4.3 Estimation of SNR shock velocities

The 5007Å/H$\beta$ ratio was used to provide a shock velocity estimate for the 12 SNR candidates where these lines are seen following Dopita et al. (1984) under the preshock physical conditions of 10 cm$^{-3}$ for density, T~9,000K and 1 $\mu$G magnetic field strength. This was done when the whole red-blue spectra were available (SAAO observations), or where only the blue spectra (with both lines) are covered (MSSSO observations). Estimates are in Table 3. Lack of [OIII] 5007Å emission in other remnants suggests shock velocities of 70 kms$^{-1}$ or less. For those remnants the [SII] lines were used for shock velocity estimates. For the SAAO spectra, covering ~3700-7000Å diagrams adapted from dop84 were used with [SII] 6731Å/H$\alpha$. Where separate blue and red spectra (MSSSO 2.3m) or where only red spectra (SAAO high dispersion) were obtained, the shock velocity was estimated using the [SII] 6717/6731Å ratio (Shull & McKee 1979) with pre-shock conditions similar to Dopita et al. (1984).

The shock velocities in Table 3 are only approximate because of the dependence of the shock optical spectrum on shock parameters which is very complex (Raymond 1979). No simple, reliable method is available and any variables are involved such as temperature, density, magnetic field strength etc. The differences in shock velocity estimates on the basis of line ratios can also be seen in the models of Cox (1972), Raymond (1979), Shull & McKee (1979) and Hartigan, Morse & Raymond (1994).

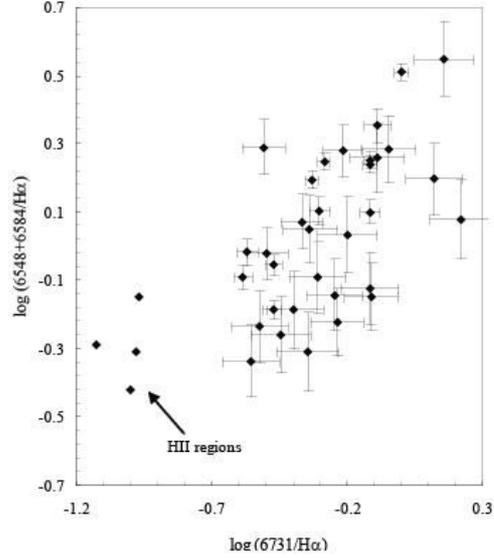

Figure 25. The relationship between [NII] 6548+6584Å/H$\alpha$ and [SII] 6731Å H$\alpha$ established by Dopita te al., (1984), showing abundance-dependent trends. Slight variations in these ratios are primarily due to metallicity changes. Every slit position for each candidate SNR from this work and line ratios from Table 3 were used in creating this diagram. The arrowed points show typical positions for HII regions taken from Hua & Llebaria (1981); Garcia-Rojas et al. (2005) and Garcia-Rojas et al. (2006). All the newly identified SNR candidates fall nicely within the region occupied by SNRs.

### 4.4 Remnant age

The observed [SII] emission line ratios can indicate whether the remnants are old. Low expansion velocity observed for old remnants have the low-density limit (Daltabuit, D'Odorico & Sabbadin 1976) while high expansion velocity objects (young remnants) have the high-density limit. The [SII] ratio for young remnants like Cassiopeia A or Kepler's SNR is typically ~0.5 (Peimbert & van der Bergh 1971; Leibowitz & Danziger 1983) while the observed [SII] ratio for this sample is between 0.93 and 1.60, typical for old remnants moving towards the low density limit.

Apart from the fragmented, faint and large angular size of most of these new optically detected entities, expected for senile, dispersing remnants, additional age confirmation arises because the [SII] ratio can be presented as a function of SNR expansion velocity (Cox 1972; Saraph & Seaton 1970; Daltabuit, D'Odorico & Sabbadin 1976). For our sample the estimated expansion velocity lies between 30 and 250 kms$^{-1}$ which is typical for evolved SNRs (for an assumed interstellar density $n_0 = 5$). However, for young remnants, this velocity is typically ~700-800 kms$^{-1}$. Hence, these new SNRs are in the snowplow or dissipation phase. Kafatos et al. (1980) state it is easier to observe an optical SNR during its isothermal (cooling) phase then in its adiabatic phase as confirmed here.

### 4.4 The Eliminating contaminants

We are confident that we have been able to identify new optical SNRs and not HII regions or evolved PNe. Our refined sample presented here is based on a combination of optical and radio morphology, environment, spectral characteristics and hunts for ionising sources. Kennicutt (2000) have shown that Galactic and Magellanic cloud HII regions have [NII]/H$\alpha$ ratios 0.6 while neither HII regions nor PNe have [SII]/H$\alpha$ > 0.6. We found no correlation of [NII]/H$\alpha$ as function of [SII] 6717/6731 Å.

Fesen, Blair & Kirshner (1985) confirmed that such a correlation does exist and derived a correlation coefficient of -0.62 bit this covers a wide range of remnant evolutionary stage. Our lack of apparent correlation is likely a selection effect as only old remnants were observed and there is no evolutionary lever arm to reveal the trend. Most probably this is the same for the sample of SNRs in M33 as reported by Smith et al. (1993), where the smallest [SII] ratio was ~1.3 or by Gordon et al. (1998), with the smallest [SII] ratio of ~1. In both case, no correlation was found.

Instead, we used Dopita et al. (1984) to show the different positions of HII regions and SNRs on a diagnostic line ratio plot of [NII]/H$\alpha$ against [SII] 6731Å/H$\alpha$ shown in Fig. 29. This plot offers a clear separation between HII regions and SNRs. The observed correlation reported by Dopita et al. (1984) is caused by metallicity variations in the inner regions of the Galaxy. According to Dopita et al. (1984) only the 6731Å line was used as collisional effects become important for this line at electron density $n_e > \sim 10^4$ m$^{-3}$ which is very rare in SNRs. What can be noticed from Fig. 20 is that all new, candidate SNRs fit in the area where known SNRs are located in this diagnostic diagram. An arrow in Fig. 29 shows where [HII] regions are grouped. Levenson et al. (1995) set up the comparable diagram (not only for [SII] 6731Å/H$\alpha$ but also [SII] (6717+6731/H$\alpha$ ) to the show position and division between these two groups, where again all 18 candidate SNRs from this work clearly fit the location of supernova remnants.

The Five-Level Atom (FIVEL) software Shaw & Dufour (1995), which computes the density and temperature for emission line spectra, was used to derive electron density $n_e$ from the ratio of the observed [SII] 6717/6731Å lines, assuming a standard temperature of 10,000 K. Of course the electron gas in SNRs can be shock heated to higher temperatures. The results for $n_e$ are given in Table 3 except where the observed values of the [SII] ratio falls near the lower density limit of the function (see Osterbrock & Ferland 2006).

## 5 Conclusions

Thanks to the high sensitivity and areal coverage of the SHS we have been able to uncover over 80 filaments and coherent nebular emissions over the Southern Galactic plane which we deem to belong to new Galactic SNR candidates based on a careful, systematic search of the SHS data. Based on our spectroscopic follow-up of about 60 objects, we have been able to confirm 18 new Galactic remnants comprising 18 new objects and 3 Galactic SNRs previously registered as candidates from existing radio observations. This work has resulted in an ~8% increase in known Galactic SNRs (Green2006). Significantly, these SNRs have been uncovered via detection of optical H$\alpha$ nebulosities and subsequent optical spectroscopic confirmation rather than from radio observations. Many more candidates await follow-up and confirmation.

Of these remnants 11 have a shell-like structure while the others are isolated filaments or emission clouds. All candidates display typical SNR optical spectra with forbidden lines of [OI] , [OII] , [O iii] , [NII] and [SII] , as well as prominent Balmer lines. All the objects have [SII]/H$\alpha > 0.5$ with all but two having this ratio $> 0.7$, confirming the presence of strong shock motions typical of SNRs and which clearly separates them from HII regions or most PNe (see Fig. 29). The high [NII]/H$\alpha$ ratio also eliminates these objects as HII regions following Kennicutt et al. (2000).

The only doubtful identification is RCW 59 (or G293.0+4.5 from the slit position), a well-known HII region where the optical structure of the shell is followed in the radio. Only one spectrum was taken of this object but it displays typical SNR characteristics. It is possible that the observed nebula is a mixture of both types of objects. More spectra and detailed radio observations are needed to make a final decision concerning the nature of RCW 59.

All candidate SNRs showed [SII] 6717/6731Å ratios typical of old Galactic SNRs. The [NII]/H$\alpha$ ratio varied between objects, as can be expected due to variations in nitrogen abundance within the Galaxy but with values usually well in excess of those of HII regions. These ratios were also found to vary inside individual filaments (see case of G315.1+2.7) in some remnants an extremely high [NII]/H$\alpha$ ratio was observed (see case of G332.5-5.6). Shock velocity estimates for all candidates were between 50 and 100 kms$^{-1}$ typical for old remnants. Typical SNR electron and pre-shock densities were also observed.

Candidates were checked against existing data from the different radio surveys, ie. the SUMSS at 843 MHz, NVSS at 1.4 GHz, Parkes at 2.4 GHz and the PMN at 4.85 GHz. Often this led to a fresh evaluation of the radio data and the recognition of clear associations not previously recognised. A check for any possible connection with an X-ray source in the vicinity was also made. Eleven objects were shown to have clear positional and feature alignment agreement with radio sources while 11 objects also revealed a possible connection with X-ray sources.

Their remain dozens of SNR candidates we have uncovered that still require spectroscopic follow-up so many additional Galactic SNRs are expected. However, we still need further optical spectra for the SNRs presented here across different parts of each remnant as well as more sensitive multi-frequency radio and X-ray observations. Finally, for the first time, we have also also optically detected about 30 filaments/diffuse emissions at the positions of known Galactic SNRs (see Stupar, Parker & Filipović 2007b) from the SHS data. According to green2006 only ~17% of known Galactic SNR had identified optical emission so new detections will shed new light on the optical spectral investigation of Galactic SNRs. We confirm the existence of a new group of SNRs which are radio quiet but optically active (see Appendix).

## 6 Acknowledgements

MS acknowledges the support of an APA PhD scholarship held at Macquarie University, Sydney, Australia. MS is also thankful to the staff of the Wide Field Astronomy Unit at the Royal Observatory Edinburgh for their help during the visual inspection of original films of the AAO/UKST H$\alpha$ Survey of the Southern Galactic Plane in August 2004. We thank our colleague David Frew for valuable discussions and for noting G281.9+8.7 with QAP during searches for large, evolved PNe. We thank the SAAO and MSSSO Time Allocation Committees for enabling the spectroscopic follow-up to be obtained. QAP and MS thank ANSTO for travel grants that enabled the spectroscopic observations to be undertaken at SAAO. We would like to thank the reviewer for constructive comments that have improved the paper.

**Appendix**

**Negative radio detection for some new remnants**

A question from this work is why only some new optically identified Galactic SNRs, based on their morphological and spectroscopic criteria, have not been detected in the available radio surveys. Blandford & Cowie (1982) state that optical emission should have associated radio emission (possibly weak), while clouds, which have been fully crossed by the radiative shocks, may be hardly detected optically. However, we raise the example of G332.5-5.6 the ``parperclip'' (see Stupar et al. 2007c), where coincident optical, radio and X-ray emission is seen. This raises another problem. The X-ray emission from a SNR must be associated with temperatures of $\sim 10^7$ K. Such high temperatures cannot be connected with cooling shock waves. The explanation is that denser regions, enveloped by the remnant and crossed with a relatively slow, cooling shock, are responsible for the optical/radio detection, while more diffuse gas remains hot and is responsible for the X-ray emission (Blandford & Cowie 1982 on the basis of McKee & Ostriker 1977). Still, the question of missing radio detection remains. We offer several possible explanations:

1. The search for radio counterparts was made with respect to existing radio survey data whose coverage is far from uniform and samples a broad range of discrete frequencies with widely varying sensitivities and resolutions. Objects may not be seen at a specific frequency, but may be detectable at other frequencies and radio telescope resolutions (see Blandford & Cowie (1982) statement about ``possibly weak'' radio recognition of remnants). A good example of this is the Vela Z SNR (see Stupar et al. 2005) which is barely on the threshold of recognition at 843 MHz, but is clearly seen at other (higher) frequencies. Also, there is the case of G315.1+2.7 (Stupar, Parker & Filipović 2007a), where a complete shell is seen at 2400 and 4800 MHz, while only one arc is vaguely seen at 843 MHz.

2. Another explanation, is that some remnants are definitely in the last dissipation phase, almost dissolved and mixed with the ISM. Whether they are truly dissolved, and what might really remain for optical detection, depends on the local ISM density and temperature. For the radio, if the magnetic field in the SNR area comes up to the level of the Galactic background magnetic field, then due to this dissipation, non thermal emission is not likely to arise and radio detection is not possible. If a light trace of non thermal radiation exists, then this might only be for small filaments. We need much better resolution radio telescopes to investigate this further such as the future Square Kilometer Array (SKA).

3. The model of Blandford & Cowie (1982) predicts the existence of a class of ``radio-quiet'' remnants which possibly lie at the lower end of the range of surface brightness (radio) at a given remnant radius. This is precisely the regime of these new remnants. This sustains our observations and our statement that we need better resolution radio telescopes as well as more sensitive optical narrow-band surveys. As an example in Stupar, Parker & Filipović (2007b) we report on the first optical H$\alpha$ detection of about 30 known Galactic SNRs recorded in the comprehensive compilation of Green (2006) and identified in the AAO/UKST H$\alpha$ survey (Parker et al. 2005). This demonstrates what is possible with new surveys with improved sensitivity, resolution and wide area coverage.

4. Our observations fit the model of Asvarov (2006) where it is clear that the process of generating new radio-emitting electrons stops with the beginning of the radiative cooling of an SNR. According to Asvarov (2006) these radio-quiet SNRs (with decreased activity in X-rays also) evolve in the homogenous ISM and should emit in the optical and HI wavelengths which is exactly what we are proposing for many of our evolved remnants.

5. One final support for our SNR identifications, especially in those cases where there are no detected X-ray sources in the vicinity, comes from the recent work of Pannuti, Schlegel & Lacey (2007).. They presented a Chandra search for X-ray counterparts of extragalactic SNRs in M 81, M 101, NGC 2403, NGC 4736 and NGC 6946. They used a sample of 138 optically identified SNRs and 50 candidate radio SNRs and found X-ray detection in only 9 optically identified SNRs and in only 12 radio SNRs that were not already known to be time-variable or associated with X-ray binaries. They argue that the role played by ambient density seriously affects searches for SNRs in those nearby galaxies at multiple wavelengths. They also conclude that the local environment in which a particular SNR is propagating plays a very important role together with other selection effects such as SN type and survey sensitivity. In another words, they found in external galaxies exactly what we have found for our Galaxy with senile remnants.